\documentclass[12pt]{article}

\usepackage{amssymb,amsfonts,amsthm}
\usepackage{amsmath}
\usepackage{bm}             
\usepackage[mathscr]{eucal}
\usepackage{cancel}
\usepackage{wasysym}

\def\be{\begin{equation}}
\def\ee{\end{equation}}
\def\bea{\begin{eqnarray}}
\def\eea{\end{eqnarray}}
\def\lb{\label}

\def\bna{\mbox{\boldmath $\nabla$}}
\def\bphi{\mbox{\boldmath $\phi$}}
\def\brho{\mbox{\boldmath $\rho$}}
\def\bsigma{\mbox{\boldmath $\sigma$}}
\def\bomega{\mbox{\boldmath $\omega$}}

\def\hsp5{\hspace{5mm}}

\theoremstyle{remark}

\newcommand{\sfrac}[2]{{\textstyle{#1\over#2}}}

\newcommand{\cH}{\mathcal{H}}
\newcommand{\cC}{\mathcal{C}}
\newcommand{\cA}{\mathcal{A}}
\newcommand{\cB}{\mathcal{B}}

\title{\sc Cosmological Perturbation Theory Revisited}

\begin{document}

\author{ \\
{\Large\sc Claes Uggla}\thanks{Electronic address:
{\tt claes.uggla@kau.se}} \\[1ex]
Department of Physics, \\
University of Karlstad, S-651 88 Karlstad, Sweden
\and \\
{\Large\sc John Wainwright}\thanks{Electronic address:
{\tt jwainwri@uwaterloo.ca}} \\[1ex]
Department of Applied Mathematics, \\
University of Waterloo,Waterloo, ON, N2L 3G1, Canada \\[2ex] }

\maketitle

\begin{abstract}

Increasingly accurate observations are driving theoretical
cosmology toward the use of more sophisticated  descriptions of
matter and the study of nonlinear perturbations of
Friedmann-Lemaitre cosmologies, whose governing equations are
notoriously complicated. Our goal in this paper is to formulate
the governing equations for linear perturbation theory in a
particularly simple and concise form in order to facilitate the
extension to nonlinear perturbations. Our approach has several
novel features. We show that the use of so-called {\it
intrinsic gauge invariants} has two advantages. It naturally
leads to: (i) a physically motivated choice of a gauge
invariant associated with the matter density, and (ii) two
distinct and complementary ways of formulating the evolution
equations for scalar perturbations, associated with the work of
Bardeen and of Kodama and Sasaki. In the first case the
perturbed Einstein tensor gives rise to a second order (in
time) linear differential operator, and in the second case to a
pair of coupled first order (in time) linear differential
operators. These operators are of fundamental importance in
cosmological perturbation theory, since they provide the
leading order terms in the governing equations for nonlinear
perturbations.

\end{abstract}

\centerline{\bigskip\noindent PACS numbers: 04.20.-q, 98.80.-k,
98.80.Bp, 98.80.Jk}

\section{Introduction}

The analysis of linear perturbations of Friedmann-Lemaitre (FL)
cosmologies was initiated by Lifshitz (1946) in a paper of
far-reaching importance. Working in the so-called synchronous
gauge, this paper showed that an arbitrary linear perturbation
can be written as the sum of three modes, a scalar mode that
describes perturbations in the matter density, a vector mode
that describes vorticity and a tensor mode that describes
gravitational waves. For many years, however, the theory was
plagued by gauge problems, {\it i.e.} by the fact that the
behaviour of the scalar mode depends significantly on the
choice of gauge. A major step in alleviating this difficulty
was taken by Bardeen (1980), who reformulated the linearized
Einstein field equations in terms of a set of {\it
gauge-invariant} variables, as an alternative to the
traditional use of the synchronous gauge. Central to Bardeen's
paper are two gauge-invariant equations that govern the
behaviour of scalar perturbations. The first of these governs
the evolution in time of a gauge-invariant gravitational ({\it
i.e.} metric) potential and the second determines a
gauge-invariant perturbation of the matter density in terms of
the spatial Laplacian of the gravitational potential. Since
this potential continues to play a central role in the study of
scalar perturbations, it seems appropriate to refer to it as
the {\it Bardeen potential}. Bardeen's paper makes clear,
however, that there is no unique way of constructing
gauge-invariant variables.

From our perspective, one drawback of Bardeen's paper is that
he performs a harmonic decomposition of the variables {\it ab
initio}, with the result that the mathematical structure of the
governing equations is somewhat obscured. In a subsequent
paper, Brandenberger, Khan and Press (1983) address this
deficiency by giving a new derivation of Bardeen's
gauge-invariant equations. They  do not perform a harmonic
decomposition, with the result that their evolution equation is
a partial differential equation rather than an ordinary
differential equation as in Bardeen's paper. However, unlike
Bardeen they restrict consideration to a spatially flat
Robertson-Walker (RW) background.\footnote{We follow the
nomenclature of Wainwright and Ellis (1997) where an FL
cosmology is a RW geometry that satisfies Einstein's field
equations.}

In subsequent developments the status of the Bardeen potential
was further enhanced by the appearance of the major review
paper by Mukhanov {\it et al} (1992), which contains a
simplified derivation of the Bardeen potential and the
evolution equation for scalar perturbations, without performing
a harmonic decomposition. However, the treatment in Mukhanov
{\it et al} (1992) is less general than that of Bardeen (1980)
and Brandenberger {\it et al} (1983) in two respects. First,
they assume the anisotropic stresses are zero, and second, they
make a specific choice of gauge invariants {\it a priori},
namely those associated with the so-called longitudinal gauge.

Currently, increasingly accurate observations are driving
theoretical cosmology towards more sophisticated models of
matter and the study of possible nonlinear deviations from FL
cosmology. Motivated by this state of affairs, our long term
goal is to provide a general but concise description of
nonlinear perturbations of FL cosmologies that will reveal the
mathematical structure of the governing equations and enable
one to make the transition between different gauge-invariant
formulations, thereby simplifying and relating the different
approaches that have been used to date.\footnote{See, for
example, Noh and Hwang (2004), Nakamura (2007) and Malik
(2007).}
In pursuing this objective we have found it necessary to
revisit linear perturbation theory, even though it is by now a
mature discipline.\footnote{For some recent reviews and books,
see, for example, Mukhanov (2005), Tsagas {\it et al} (2008),
Weinberg (2008), Durrer (2008), Malik and Wands (2009) and Lyth
and Liddle (2009).} Our intent in the present paper is to
formulate the governing equations for the linear theory in a
particularly simple and concise form in order to facilitate the
extension to nonlinear perturbations.

Based on earlier work by Bruni {\it et al} (1997) on
gauge-invariant higher order perturbation theory, Nakamura
(2003) introduced a geometrical method for constructing gauge
invariants for linear and nonlinear (second order)
perturbations which he later applied to derive the governing
equations (see Nakamura (2006) and Nakamura (2007)). In the
present paper we use a dimensionless version of Nakamura's
method for constructing gauge invariants, but we complement it
with the observation that gauge invariants are of two distinct
types: {\it intrinsic gauge invariants}, i.e., gauge invariants
that can be constructed from a given tensor alone, and {\it
hybrid gauge invariants}, {\it i.e.} gauge invariants that are
constructed from more than one tensor.

In Nakamura's approach, the linear perturbation of any tensor
is written as the sum of a gauge-invariant quantity and a
gauge-variant quantity, which is the Lie derivative of the zero
order tensor with respect to a suitably chosen vector field
$X$. A choice of $X$ yields a set of gauge-invariant variables
that are associated with a specific fully fixed gauge. We will
show that for the metric tensor there exist two natural
complementary choices of $X$ that yield intrinsic metric gauge
invariants. One choice, used in all of Nakamura's papers, leads
to the two gauge-invariant metric potentials of Bardeen (1980),
which are associated with the so-called {\it Poisson
gauge}.\footnote{The Poisson gauge, which was introduced by
Bertschinger (1996) (see his equation (4.46)), is a
generalization of the longitudinal gauge, which only applies to
scalar perturbations. } The other choice leads to the two
gauge-invariant metric potentials of Kodama and Sasaki (1984),
which are associated with the so-called {\it uniform curvature
gauge}.\footnote{ See, for example, Malik and Wands (2009),
page 20, and other references given there.} We will show that
these two preferred choices lead to two distinct ways in which
to present the linearized Einstein field equations: with the
Bardeen choice the evolution of linear scalar perturbations is
governed by a {\it second order} (in time) linear partial
differential operator, while with the Kodama-Sasaki choice the
evolution is governed by two coupled temporal {\it first order}
linear operators.

The plan of the paper is as follows. In
Section~\ref{sec:geomdef} we discuss the geometrical
construction of gauge-invariants: we focus on the metric tensor
and, with the Einstein tensor and the stress-energy tensor in
mind, on mixed rank two tensors. In Section~\ref{sec:lin} we
use intrinsic gauge invariants to derive the general governing
equations for linear perturbations in two gauge-invariant forms
associated with the Poisson and the uniform curvature gauges.
The required expressions for the Einstein gauge invariants are
derived efficiently in Appendix~\ref{app:curv}, where we also
give a general concise formula that expresses the Riemann gauge
invariants in terms of the metric gauge invariants. One of the
ingredients in our derivation is the so-called Replacement
Principle, which is formulated in Appendix~\ref{app:repl}. In
Section~\ref{sec:interpret} we give an interpretation of the
intrinsic matter gauge invariants and specialize our equations
to the cases of a perfect fluid and a scalar field.
Section~\ref{sec:discussion} contains a brief discussion of
future developments.

\section{Geometrical definition of
gauge invariants}\label{sec:geomdef}

\subsection{General formulation} \label{sec:genform}

Following standard cosmological perturbation theory (see for example,
Chapter 7.5 in Wald (1984)), we consider a 1-parameter family of
spacetimes $g_{ab}(\epsilon)$, where $g_{ab}(0)$, the
unperturbed metric, is a RW metric, and $\epsilon$ is referred
to as the {\it perturbation parameter}.\footnote{We use Latin
letters $a,b,\dots,f$ to denote abstract spacetime indices.}
We assign physical dimension $length$ to the scale factor
$a$ of the RW metric and $(length)^2$ to $g_{ab}(\epsilon)$. Then
the conformal transformation
\be\label{bar_g}  g_{ab}(\epsilon) = a^2{\bar g}_{ab}(\epsilon),\ee
yields a dimensionless metric ${\bar g}_{ab}(\epsilon)$. Our
reason for making this choice\footnote{An alternative choice in
cosmology is to make $a$ dimensionless and let the spacetime
coordinates of ${\bar g}_{ab}(0)$ have dimension $length$ (see,
for example, Malik and Wands (2009), page 48). This choice is
unsuitable for our purposes since it does not lead naturally to
perturbative equations involving dimensionless quantities.}
concerning the allocation of physical dimensions is that it
enables one to create dimensionless quantities by multiplying
by the appropriate power of $a$, leading to simple perturbation
equations that do not contain $a$ explicitly. We refer to
Appendix~\ref{app:curv}, where this process is applied.

The Riemann tensor associated with the metric
$g_{ab}(\epsilon)$ is a function of $\epsilon$, denoted
$R^{ab}\!_{cd}(\epsilon)$, as is the Einstein tensor,
$G^a\!_b(\epsilon)$. The stress-energy tensor of the matter
distribution is also be assumed to be a function of $\epsilon$,
denoted $T^a\!_b(\epsilon)$. We include all these possibilities
by considering a 1-parameter family of tensor fields
$A(\epsilon)$, which we assume can be expanded in powers of
$\epsilon$, {\it i.e.} as a Taylor series:
\be\label{taylor} \mathrm{A}(\epsilon) = {}^{(0)}\!\mathrm{A} +
\epsilon\,{}^{(1)}\!\mathrm{A} + \sfrac{1}{2}\epsilon^2\,
{}^{(2)}\!\mathrm{A} + \dots\, .\ee
The coefficients are given by\footnote{The notation $A(\epsilon)$
should be viewed as shorthand for $A(x,\epsilon)$, indicating
that the tensor fields are functions of the spacetime coordinates,
which necessitates the use of partial differentiation with respect
to $\epsilon$.}
\be\label{perturb}  {}^{(0)}\!\mathrm{A} = A(0),  \qquad
{}^{(1)}\!\mathrm{A}= \left.\frac{\partial
\mathrm{A}}{\partial\epsilon}\right|_{\epsilon=0}, \qquad
{}^{(2)}\!\mathrm{A}= \left.\frac{\partial^2
\mathrm{A}}{\partial\epsilon^2}\right|_{\epsilon=0},  \quad
\dots, \ee
where  ${}^{(0)}\!A$ is called the {\it unperturbed value},
${}^{(1)}\!\mathrm{A}$ is called  the {\it first order (linear)
perturbation} and ${}^{(2)}\!\mathrm{A}$ is called  the {\it
second order perturbation} of $A(\epsilon)$.

The primary difficulty in cosmological perturbation theory is
that the perturbations of a tensor field $A(\epsilon)$ depend
on the choice of gauge, and hence cannot be directly related to
observations. It is therefore desirable to formulate the theory
in terms of {\it gauge-invariant} quantities, {\it i.e.} to
replace the gauge-variant perturbations ${}^{(1)}\!\mathrm{A},
{}^{(2)}\!\mathrm{A}, \dots$ of $A(\epsilon)$ by
gauge-invariant quantities. In this paper we restrict our
attention to first order, {\it i.e.} linear, perturbations, but
with a view to subsequently working with higher order
perturbations we use a method pioneered by Nakamura
(2003), and adapt it so as to create quantities that are
gauge-invariant \emph{and} dimensionless.

A linear gauge transformation is represented in coordinates
by the equation
\be {\tilde x}^a =  x^a  + \epsilon \xi ^a + \dots, \ee
where $\xi^a$ is an arbitrary dimensionless vector
field on the background.
Given a family of tensor fields $A(\epsilon)$
 the change induced in the first order
perturbation ${}^{(1)}\!A$ by a gauge transformation
is determined by
\be \label{delta_A} \Delta {}^{(1)}\!A =
\pounds_{\xi}{}^{(0)}\!A,  \ee
where $\pounds_{\xi}$ denotes the Lie derivative with respect
to $\xi^a$ and $\Delta {}^{(1)}\!A := {}^{(1)}\!{\tilde A} -
{}^{(1)}\!A$ (see, for example, Bruni {\it et al} (1997),
equations (1.1) and (1.2)). We now introduce an as yet
arbitrary dimensionless vector field $X$ on the background
which we use to define the dimensionless object\footnote{When
we consider second order perturbations in a subsequent paper,
we will denote $\xi^a$ and $X^a$ by $^{(1)}\!\xi^a$ and
$^{(1)}\!X^a$, and introduce a second pair of vector fields
denoted $^{(2)}\!\xi^a$ and $^{(2)}\!X^a$.}
\be\label{bold_A} {}^{(1)}\!{\bf A}[X] := a^n \left({}^{(1)}\!A
- \pounds_{X} {}^{(0)}\!A \right),  \ee
where we assume that $A(\epsilon)$ is such that $a^n
A(\epsilon)$ is dimensionless. It follows from~\eqref{delta_A}
and~\eqref{bold_A} that
\be\label{delta_A2} \Delta {}^{(1)}\!{\bf A}[X] = a^n \left(
\pounds_{\xi}{}^{(0)}\!A - \pounds_{\Delta X}{}^{(0)}\!A\right)
= a^n \pounds_{\xi-\Delta X} {}^{(0)}\!A . \ee
The key step is to choose an $X$  that satisfies
\be \label{delta_X} \Delta X^a = \xi^a, \ee
under a gauge transformation. With this choice,
\eqref{delta_A2} implies that $\Delta {{}^{(1)}\!\bf A}[X] =
0$, {\it i.e.}, ${{}^{(1)}\!\bf A}[X]$ is gauge-invariant. We
say that ${{}^{(1)}\!\bf A}[X]$ is the {\it gauge invariant
associated with ${}^{(1)}\! A$ by $X$-compensation}.
Equations~\eqref{delta_A}, \eqref{bold_A} and~\eqref{delta_X}
are central to our version of Nakamura's method for
constructing gauge invariants associated with the first order
perturbation of a tensor $A$ (see Nakamura (2007), equations
(2.19), (2.23) and (2.26)). In what follows we will drop the
superscript $^{(1)}$ on ${\bf A}$ for convenience since in this
paper we are dealing only with first order perturbations.

The above `gauge compensating vector field' $X$, which for
brevity we shall refer to as the {\it gauge field}, requires
comment. Unlike the geometric and matter tensor fields such as
$g_{ab}(\epsilon)$ and $T^a\!_b(\epsilon)$ it is not the
perturbation of a corresponding quantity on the background
spacetime. Instead it should be viewed as a vector field on the
background spacetime that is constructed from the linear
perturbations of the geometric and matter tensors in such a way
that~\eqref{delta_X} holds. We will construct specific examples
of $X$ in section~\ref{sec:metric}. We note that in choosing
the gauge field $X$ we are essentially fixing the gauge ({\it i.e.}
making a choice of gauge), which is accomplished in the
traditional approach by making a choice of the vector field
$\xi$ that determines the gauge transformation.\footnote
{See, for example, Malik and Wands (2009); equations (6.17),
(7.3) and (7.4) provide an example in connection with the
metric tensor.} One advantage of using the gauge
field $X$ is that one immediately obtains a geometric connection
between the gauge invariants associated with different choices
of gauge. This matter is discussed in more detail in
Uggla and Wainwright (2011).

Before continuing we briefly digress to point out that
gauge invariants associated with a tensor $A$  are of two
distinct types: those that are solely constructed from
components of ${}^{(1)}\!A$ and ${}^{(0)}\!A$ are called {\it
intrinsic} gauge invariants, while those that depend on the
components of another perturbed tensor are called {\it hybrid}
gauge invariants. In particular if the gauge field $X$ is
formed solely from components of ${}^{(1)}\!A$ and
${}^{(0)}\!A$, then ${\bf A}[X]$ is an intrinsic gauge
invariant, otherwise ${\bf A}[X]$ is a hybrid gauge invariant.

In the following sections we will calculate the quantities in
equations~\eqref{delta_A} and~\eqref{bold_A} for various
geometric objects $A$. To do this it is necessary to use the
well known formulae for the Lie derivative. The formula for a
tensor of type $(1,1)$, which we now give, establishes the
pattern:
\be \label{lie_partial} \pounds_{\xi}A^a\!_b  =
A^a\!_{b,c}{}\xi^c + \xi^c\!_{,b} A^a\!_c - \xi
^a\!_{,c}A^c\!_b, \ee
where $_,$ denotes partial differentiation. In a formula such
as \eqref{lie_partial} one can replace the partial derivatives
by covariant derivatives. For our purposes it is convenient to
use the covariant derivative ${}^0\!\bar{\bna}\!_a $ associated
with the {\it unperturbed conformal metric $\bar{g}_{ab}(0)$}:
\be \label{lie_cov} \pounds_{\xi}A^a\!_b  =
({}^0\!\bar{\bna}\!_c A^a\!_b)\xi^c +
({}^0\!\bar{\bna}\!_b\xi^c) A^a\!_c - ({}^0\!\bar{\bna}\!_c\xi
^a) A^c\!_b. \ee

We also need to work in a coordinate frame so that we can
calculate time and spatial components separately. We thus
introduce local coordinates\footnote{We use Greek letters to
denote spacetime coordinate indices on the few occasions that
they occur, and we use Latin letters $i,j,k,m$ to denote
spatial coordinate indices, which are lowered and raised using
$\gamma_{ij}$ and its inverse  $\gamma^{ij}$, respectively.}
$x^\mu=(\eta,x^i)$, with $\eta$ being the usual conformal time
coordinate\footnote{Since we assigned $a$ to have physical
dimension $length$, the conformal time $\eta$ and the
conformal spatial line-element $\gamma_{ij}dx^idx^j$ are
dimensionless. We choose the $x^i$ to be dimensionless,
which implies that the $\gamma_{ij}$ are also dimensionless.}
for the RW metric $g_{ab}(0)$, and such that the
unperturbed conformal metric $\gamma_{ab} := \bar{g}_{ab}(0)$
has components
\be \gamma_{00}=-1\, ,\qquad \gamma_{0i}=0\, ,\qquad
\gamma_{ij}\, ,\ee
where $\gamma_{ij}$ is the metric of a spatial geometry of
constant curvature. The curvature index of the RW metric,
denoted $K$, determines the sign of the curvature of
the spatial geometry, and if non-zero can be scaled to be
$+1$ or $-1$ (see, for example, Plebanski and Krasinski
(2006), page 261).

 The spacetime covariant derivative
${}^0\!\bar{\bna}\!_a $ determines a temporal derivative
${}^0\!\bar{\bna}\!_0 \, A = \partial_\eta A$, where
$\partial_\eta$ denotes partial differentiation with respect to
$\eta$, and a spatial covariant derivative
${}^0\!\bar{\bna}\!_i $ that is associated with the spatial
metric $\gamma_{ij}$. We introduce the notation
\be\label{bold_D}  {\bf D}_i A: = {}^0\!\bar{\bna}\!_i A. \ee
The derivative operators $\partial_\eta$ and ${\bf D}_i$ will
be used throughout this paper once local coordinates have been
introduced. However, for simplicity we shall denote
the derivative of a function $f(\eta)$ that depends only
on $\eta$ by $f'(\eta)$.

 With our present allocation of dimensions,
the scalar $\cH$ defined by
\be\label{hubble} {\cal H} := \frac{a'}{a} = aH, \ee
where $H$ is the Hubble scalar,\footnote {Recall that $H :=
\frac{1}{a}\frac{da}{dt}$, where $t$ is cosmic time, and that
$\frac{dt}{d\eta} = a$.} is dimensionless. We shall refer to it
as the {\it dimensionless Hubble scalar}.  The use of this
scalar, e.g. by Mukhanov {\it et al} (1992) (see page 218), is
essential in eliminating $a$ from the perturbation equations.

\subsection{Metric gauge invariants} \label{sec:metric}

We expand ${\bar g}_{ab}(\epsilon)$, defined by
equation~\eqref{bar_g}, in powers of $\epsilon$:
\begin{equation*}  {\bar g}_{ab}(\epsilon) =
{}^{(0)} {\bar g}_{ab} + \epsilon\,  {}^{(1)} {\bar g}_{ab} + \dots\, ,
\end{equation*}
and label the unperturbed metric and (linear) metric perturbation according to
\be\label{gamma,f}  \gamma_{ab} := {}^{(0)} {\bar g}_{ab} =
{\bar g}_{ab}(0), \qquad  f_{ab} := {}^{(1)} {\bar g}_{ab} =
\frac{\partial{\bar g}_{ab}}{\partial\epsilon}(0), \ee
which is consistent with~\eqref{perturb}. Applying the general
transformation law~\eqref{delta_A} to the metric tensor
$g_{ab}(\epsilon) = a^2{\bar g}_{ab}(\epsilon)$ we obtain
\be\label{delta_fab} \Delta ^{(1)}\!g_{ab} =
\pounds_\xi{}^{(0)}\! g_{ab},\quad \text{or,
equivalently}, \quad \Delta f_{ab} = a^{-2}\pounds_{\xi}
(a^2\gamma_{ab}), \ee
in terms of the notation~\eqref{gamma,f}. The gauge invariant
${\bf f}_{ab}[X]$ associated with the metric perturbation
$f_{ab}$ by $X$-compensation, given by~\eqref{bold_A}, assumes
the form
\be \label{bf_f}  {\bf f}_{ab}[X] = f_{ab} -
a^{-2}{\pounds}_{X}(a^2\gamma_{ab}). \ee
Introducing local coordinates and using~\eqref{lie_partial}
and~\eqref{lie_cov} adapted to a $(0,2)$ tensor,
equations~\eqref{delta_fab} and~\eqref{bf_f} lead to
\begin{subequations}\label{delta_f}
\begin{xalignat}{2}
\Delta {f}_{00} &= -2(\partial_\eta + {\cal H}){\xi}^{0},
&\quad {\bf f}_{00}[X] &= f_{00} + 2(\partial_\eta + {\cal H})X^{0},\\
\Delta {f}_{0 i} &= -{\bf D}_i{\xi}^0 +\partial_\eta {\xi}_i,
&\quad {\bf f}_{0 i}[X] &= f_{0 i}  + {\bf D}_i X^0 - \partial_\eta X_i ,\\
\Delta {f}_{ij} &= 2{\cal H}\,\xi^0\gamma_{ij} + 2{\bf
D}_{(i}\xi_{j)}, &\quad {\bf f}_{ij}[X] &= f_{ij} - 2{\cal
H}X^0\gamma_{ij} - 2{\bf D}_{(i}X_{j)}.
\end{xalignat}
\end{subequations}

In order to construct a gauge field $X$ that
satisfies~\eqref{delta_X}, {\it using only the metric}, we need
to decompose the metric perturbation $f_{ab}$ into scalar,
vector and tensor modes.\footnote{In order to guarantee that
the functions $B, B_i, C, C_i$ and $ C_{ij}$ in \eqref{split_f}
are uniquely determined by $f_{0i}$ and $f_{ij}$ we need to
assume that the inverses of ${\bf D}^2$, ${\bf D}^2 + 2K$ and
${\bf D}^2 + 3K$ exist. See the proposition in Appendix
\ref{app:prop}. See also Nakamura (2007), following equation
(4.15), for a helpful discussion of this matter.} We introduce
the notation\footnote{We are denoting the scalar mode functions
by $\varphi, B, C$ and $\psi$, in agreement with Mukhanov {\it
et al} (1992) (see equation (2.10), but note the different
signature) and Malik and Wands (2009) (see equations
(2.7)-(2.12)), with the difference that we use $C$ instead of
$E$. Bardeen (1980) used the notation $A, -B, H_T$ and $-H_L
+\sfrac13 {\bf D}^2 H_T$ for these functions, the choice of the
fourth one being motivated by harmonic decomposition.
Bardeen's notation has been used by subsequent authors, for
example, Kodama and Sasaki (1984) and Durrer (1994), although the
latter author replaced $-B$ by $B$.}
\begin{subequations}\label{split_f}
\begin{align}
f_{00} &= -2\varphi, \\
f_{0i} &= {\bf D}_i B + B_i,\\
f_{ij} &= -2\psi \gamma_{ij} + 2{\bf D}_i {\bf D}_j C + 2{\bf D}_{(i} C_{j)} + 2C_{ij},
\end{align}
\end{subequations}
where the vectors $B_i$ and $C_i$ and the tensor $C_{ij}$
satisfy
\begin{equation*}
 {\bf D}^i B_i = 0, \qquad  {\bf D}^i C_i = 0, \qquad
C^i\!_i = 0, \qquad   {\bf D}^i C_{ij} = 0.
\end{equation*}
The vector $\xi$ is also decomposed into a scalar mode and a
vector mode with components
\be \xi^0, \qquad  \xi^i = {\bf D}^i \xi + \tilde {\xi}^i.
\label{split_xi} \ee
It follows from~\eqref{delta_f}, \eqref{split_f}
and~\eqref{split_xi} that
\begin{subequations}\label{delta_split_f}
\begin{alignat}{2}
\Delta \varphi &= (\partial_\eta + {\cal H})\xi^0 ,&\qquad
\Delta B &= -\xi^0 + \partial_\eta\xi ,\qquad \Delta C = \xi
,\qquad \Delta \psi = -{\cal H}\xi^0 , \label{scalar}\\
\Delta B_i &= \partial_\eta\tilde{\xi}_i ,&\qquad \Delta C_i &= \tilde{\xi}_i , \label{vector}  \\
\Delta C_{ij} &= 0. && \label{tensor}
\end{alignat}
\end{subequations}

We can draw two immediate conclusions. First, it follows
from~\eqref{vector} and~\eqref{tensor} that $B_i - C_i^\prime$
and $C_{ij}$ are gauge invariants. We introduce the following
bold-face notation:
\be\label{boldB,C} {\bf B}_i :=  B_i - \partial_\eta C_i,
\qquad   {\bf C}_{ij} := C_{ij}. \ee
Second, by inspection of~\eqref{split_xi}, \eqref{scalar}
and~\eqref{vector} we obtain
\be\label{delta_special}   \Delta( {\bf D}_i C + C_i)  = \xi_i,
\qquad  \Delta\chi = \Delta\left(\frac{\psi}{\cal H}\right) = -
\xi^0, \ee
where we have introduced the notation
\be\label{chi} \chi :=  B - \partial_\eta C .  \ee
We are now in a position to satisfy the
requirement~\eqref{delta_X}. Firstly, referring
to~\eqref{delta_special}, we can satisfy the {\it spatial part}
$\Delta X^i = \xi^i$ of the requirement by choosing
\be\label{X_i}  X_i = {\bf D}_i C + C_i, \ee
which we will take to be our default choice  for $X_i$.
With this choice, the expressions~\eqref{delta_f}  for the
components of the gauge invariant ${\bf f}_{ab}[X]$, when
combined with~\eqref{split_f}, assume the form
\begin{subequations} \label{bold_f_split}
\begin{align}
{\bf f}_{00}[X] &= -2  \Phi[X] \, ,\\
{\bf f}_{0 i}[X] &= {\bf D}_i {\bf B}[X] + {\bf B}_i\, ,\\
{\bf f}_{ij}[X] &= -2\Psi[X] \gamma_{ij} + 2{\bf C}_{ij}\, .
\end{align}
where
\be\label{generic_scalar} \Phi[X] := \varphi - (\partial_\eta +
{\cal H})X^{0},  \qquad \Psi[X]  := \psi + {\cal H}X^0, \qquad
{\bf B}[X] := \chi + X^0 , \ee
\end{subequations}
and ${\bf B}_i, {\bf C}_{ij}$ and $\chi$ are given
by \eqref{boldB,C} and \eqref{chi}, respectively.

Secondly, referring to~\eqref{delta_special}, we can satisfy
the {\it timelike part}  $\Delta X^0 = \xi^0$ of the
requirement \eqref{delta_X} in two obvious ways, by choosing
\be\label{X_Poisson} X^0 =  X_{\mathrm{p}}^0 := - \chi, \quad
\text{or} \quad  X^0 =  X_{\mathrm{c}}^0 := - \frac{\psi}{\cal H}, \ee
which leads to the metric gauge invariants associated with the
{\it Poisson gauge}, or the  {\it uniform curvature gauge},
respectively. On substituting these choices
into~\eqref{generic_scalar} we obtain the conditions
\be  \label{choices}  {\bf B}[ X_{\mathrm{p}}] = 0 \quad
\text{and} \quad \Psi[ X_{\mathrm{c}}] = 0,  \ee
which characterize these two gauge choices.

\subsubsection*{The Poisson gauge invariants}

On substituting the first of equations~\eqref{X_Poisson}
into~\eqref{bold_f_split} we obtain
\be\label{bf_Poisson} {\bf f}_{00}[ X_{\mathrm{p}}] := -2\Phi\, ,\qquad
{\bf f}_{0i}[ X_{\mathrm{p}}] := {\bf B}_i\, ,\qquad
{\bf f}_{ij}[ X_{\mathrm{p}}] := -2\Psi\gamma_{ij}
+ 2{\bf C}_{ij}\, , \ee
where
\be \label{poisson_scalar}   \Phi := \Phi[ X_{\mathrm{p}}] =
\varphi +\left(\partial_\eta + {\cal H} \right)\chi, \qquad
\Psi := \Psi[ X_{\mathrm{p}}] = \psi - {\cal H}\chi . \ee
Here $\Phi$ and $\Psi$ are the scalar metric gauge invariants
associated with the Poisson gauge,\footnote{The gauge-fixing
conditions for the Poisson gauge are $B =C = 0, C_i = 0$ in
\eqref{split_f}.} and $\Psi$ is the Bardeen potential.

\subsubsection*{The uniform curvature gauge invariants}

On substituting the second of equations~\eqref{X_Poisson}
into~\eqref{bold_f_split} we obtain
\be\label{bf_curv} {\bf f}_{00}[ X_{\mathrm{c}}] = -2{\bf A} ,\qquad
{\bf f}_{0i}[ X_{\mathrm{c}}] = {\bf D}_i {\bf B} +  {\bf B}_i , \qquad {\bf
f}_{ij}[ X_{\mathrm{c}}] = 2{\bf C}_{ij} , \ee
where
\be\label{curv_scalar} {\bf A} := \Phi[ X_{\mathrm{c}}] = \varphi +
\left(\partial_\eta + {\cal H} \right) \frac{\psi}{\cal H},
\qquad {\bf B} := {\bf B}[ X_{\mathrm{c}}] = \chi - \frac{\psi}{\cal H}. \ee
Here ${\bf A}$ and ${\bf B} $ are the scalar metric gauge
invariants associated with the uniform curvature
gauge,\footnote{The gauge-fixing conditions for the uniform
curvature gauge are $\psi = C = 0, C_i = 0$ in
\eqref{split_f}.} introduced by Kodama and Sasaki
(1984).\footnote{See equations (3.4) and (3.5), noting that
$H_L + n^{-1}H_T \equiv -\psi$ and $B - k^{-1}H_T' \equiv
\chi$.}

In concluding this section we note that the gauge fields $X$
used to construct the above gauge invariants have the same
spatial components $X^i$ given by~\eqref{X_i} in both cases,
leading to~\eqref{bold_f_split}, with the vector and tensor
modes described by the gauge invariants ${\bf B}_i $ and ${\bf
C}_{ij}$, respectively. The difference lies in the scalar
metric gauge invariants which are related according
to\footnote{These relation have recently been given by
Christopherson {\it et al} (2011). See their equations (4.22)
and (4.23).}
\be\label{curv_Poisson} {\bf A} = \Phi + \left(\partial_\eta +
{\cal H} \right) \frac{\Psi}{\cal H} , \qquad  {\bf B} = -
\frac{\Psi}{\cal H}, \ee
as follows from \eqref{poisson_scalar} and \eqref{curv_scalar}.
In both cases the gauge invariants are {\it intrinsic} since
the gauge field $X$ depends only on the metric.

A reader of this paper should be aware of the lack of agreement
in the literature on labelling the scalar metric gauge
invariants associated with the Poisson gauge. Our choice of
$(\Phi, \Psi)$ in~\eqref{poisson_scalar} is the one initiated
by Mukhanov {\it et al} (1992), and subsequently used by
Nakamura (see, for example, Nakamura (2006)) and Malik and
Wands (2009). On the other hand Durrer (2008) and Liddle and
Lyth (2000) reverse the roles and use $(\Psi, \Phi)$, while
Kodama and Sasaki (1984) use $(\Psi, - \Phi)$. Bardeen's
original notation is $(\Phi_A, - \Phi_H)$.

\subsection{Gauge invariants for mixed rank 2 tensors}

In this subsection we consider a rank two tensor $A^a\!_b$,
such that $A_{ab}$ is symmetric and  $a^2 A^a\!_b$ is
dimensionless. We expand $A^a\!_b$ in a Taylor series in
$\epsilon$ as in~\eqref{taylor}, and assume that
${}^{(0)}\!A^a\!_b$ obeys the background symmetries, which
means it is spatially homogeneous and isotropic:
\be\label{A_0} {\bf D}_i{}^{(0)}\!A^\alpha\!_\beta = 0, \qquad
{}^{(0)}\!A^0\!_i = {}^{(0)}\!A^i\!_0 = 0 ,\qquad
{}^{(0)}\!A^i\!_j = \sfrac13\,\delta^i\!_j\,{}^{(0)}\!A^k\!_k .
\ee
We introduce the notation
\be\label{script_AC} {\cal A}_A :=  a^2( -{}^{(0)}\!
A^0\!_0 + \sfrac13 {}^{(0)}\!A^k\!_k), \qquad {\cal C}_A^2 := -
\frac{ ({}^{(0)}A^k\!_k)^\prime }{3({}^{(0)}A^0\!_0)^\prime},
\ee
where as before $'$ denotes differentiation with respect to
$\eta$. We further assume that $A^a\!_b$ satisfies the
conservation law $\bna\!_aA^a\!_b=0$. It follows that in the
background
\be  \label{cons} a^2({}^{(0)}\!A^0\!_0)^\prime =
3a^2\cH ( -{}^{(0)}\! A^0\!_0 +
\sfrac13 {}^{(0)}\!A^k\!_k) =  3{\cal H}{\cal A}_A, \ee
which, in conjunction with~\eqref{script_AC}, implies that
\be \label{cal_A'}  {\cal A}_A^\prime = -  (1 + 3{\cal
C}_A^2){\cal H}{\cal A}_A.  \ee

We can now calculate the gauge invariants $ {\bf A}^a\!_b[X]$
associated with $^{(1)}\!A^a\!_b$ by $X$-compensation, as
defined by equation~\eqref{bold_A} with $n=2$. It is convenient
to decompose $^{(1)}\!A^i\!_j$ into its trace $^{(1)}\!A^k\!_k$
and tracefree part defined by
\be {}^{(1)}\!{\hat A}^i\!_j := {}^{(1)}\!A^i\!_j - \sfrac13
{}^{(1)}\!A^k\!_k \,\delta^i\!_j .  \ee
A straightforward calculation using~\eqref{bold_A},
\eqref{lie_partial}, \eqref{lie_cov} and~\eqref{A_0} leads
to\footnote{We do not include the ${}^{(1)}\!A^i\!_0$
components since they can be expressed in terms of the other
components and the metric perturbation, due to the assumed
symmetry.}
\begin{subequations} \label{DeltaAtens}
\begin{align}
{\bf A}^0\!_0[X] &=a^2\, {}^{(1)}\!A^0\!_0 - 3\cH \cA_A X^0 \\
{\bf A}^0\!_i[X] & = a^2\,{}^{(1)}\!A^0\!_i
+ {\cal A}_A{\bf D}_i X^0, \label{AiX} \\
{\bf A}^k\!_k[X] &=
a^2\, {}^{(1)}\!A^k\!_k + 9\cH \cA_A \cC_A^2X^0, \\
{\hat{\bf A}}^i\!_j[X]
&= a^2\,{}^{(1)}\!{\hat A}^i\!_j . \label{hatA,X}
\end{align}
\end{subequations}
In deriving these equation we have used~\eqref{script_AC} and~\eqref{cons}
to express $ {}^{(0)}\!A^0\!_0,  {}^{(0)}\!A^k\!_k$ and
their derivatives in terms of $\cA_A$ and $\cC_A^2$.

Equation~\eqref{hatA,X} implies that ${\hat{\bf A}}{}^i\!_j[X]$
is an {\it intrinsic} gauge invariant since it is constructed
solely from the components of ${}^{(1)}\!A^a\!_b$. We denote
this quantity by
\be  \label{hatA} {\hat{\bf A}}^i\!_j := {\hat{\bf A}}^i\!_j[X]
= a^2\,{}^{(1)}\!{\hat A}^i\!_j .  \ee
One can form two additional intrinsic gauge invariants
by taking suitable combinations of  ${\bf A}^0\!_0
[X], {\bf A}^0\!_i [X]$ and ${\bf A}^k\!_k [X]$. Indeed it
follows from~\eqref{DeltaAtens} that
\begin{subequations}
\begin{align}
\label{bf_A}  {\bf A} &:= {\cal C}_A^2 {\bf A}^0\!_0 [X] +
\sfrac13 {\bf A}^k\!_k [X] = a^2({\cal C}_A^2 {}^{(1)}\!A^0\!_0
+ \sfrac13 {}^{(1)}\!A^k\!_k),\\
\label{bf_Ai}  {\bf A}_i &:= -\left( {\bf D}_i {\bf
A}^0\!_0 [X] + 3 {\cal H}{\bf A}^0\!_i [X]\right) = -
a^2\left( {\bf D}_i {}^{(1)}\!A^0\!_0 + 3 {\cal H}
{}^{(1)}\!A^0\!_i \right),
\end{align}
\end{subequations}
which implies that ${\bf A}$ and ${\bf A}_i$ are {\it intrinsic
gauge-invariants.}

In summary, the tensor $A^a\!_b$ can be described by the three
intrinsic gauge invariants ${\hat{\bf A}}{}^i\!_j$, ${\bf A}$,
and ${\bf A}_i$, given by~\eqref{hatA}, \eqref{bf_A}
and~\eqref{bf_Ai}, and one hybrid gauge invariant ${\bf
A}^0\!_i[X] $, given by~\eqref{AiX}. In section
\ref{sec:genform2} we will use these objects, constructed in
terms of the Einstein tensor and the stress-energy tensor, to
give a concise derivation of the governing equations in
gauge-invariant form for linear perturbations of FL.

\section{Linearized governing equations}\label{sec:lin}

\subsection{General formulation}\label{sec:genform2}

In this section we work with the linear perturbations of the
Einstein tensor and the stress-energy tensor, denoted by $
{}^{(1)}G^a\!_b$ and ${}^{(1)}T^a\!_b$, and defined via
equation~\eqref{perturb}. The corresponding unperturbed
quantities are labelled by a superscript $^{(0)}$.

We begin by imposing the background Einstein equations
${}^{(0)}\!G^a\!_b = {}^{(0)}\!T^a\!_b$. The non-zero
components are given by\footnote{See, for example,
Mukhanov {\it et al} (1992), equation (4.2), noting the difference in signature.}
\begin{subequations}\label{GT0}
\begin{align}
a^2\,{}^{(0)}\!G^0\!_0 &= -3({\cal H}^2 + K) \hskip-0.8cm
&=&\,\, - a^2{}^{(0)}\!\rho\, \hskip-3.2cm &=&\,\,\, a^2\,{}^{(0)}\!T^0\!_0,\\
a^2\, {}^{(0)}\!G^i\!_j &= -(2{\cal H}^\prime + {\cal H}^2 +
K)\delta^i\!_j \hskip-2.0cm &=& \,\,\, a^2{}^{(0)}\!p\,\delta^i\!_j\, \hskip-2cm
&=&\,\,\, a^2{}^{(0)}\!T^i\!_j,
\end{align}
\end{subequations}
where ${\cal H}$ is given by~\eqref{hubble} and
$K$ is the curvature index. It follows
from~\eqref{GT0}, \eqref{script_AC} and~\eqref{cons}, with $A$
replaced by $G$ and $T$, respectively, that
\begin{subequations}\label{ATCT}
\begin{xalignat}{2}
{\cal A}_G &= 2(-{\cal H}^\prime + {\cal H}^2 + K), &\quad {\cal
A}_T &=  a^2 ({}^{(0)}\!\rho + {}^{(0)}\!p),\\
{\cal A}_G^\prime &= -  (1 + 3{\cal C}_G^2){\cal H}{\cal A}_G,
&\quad {\cal C}_T^2 &= \frac{{}^{(0)}\!p^\prime
}{{}^{(0)}\!\rho^\prime}. \label{CT}
\end{xalignat}
\end{subequations}
The conservation law~\eqref{cons}, with $A$ replaced
by $T$, gives
\be\label{cons_T} a^2({}^{(0)}\!\rho)^\prime = - 3{\cal
H}{\cal A}_T = -3\cH a^2({}^{(0)}\!\rho + {}^{(0)}\!p). \ee
The background Einstein equations imply that $\cA_G = \cA_T$
and ${\cal C}_G^2 = {\cal C}_T^2$. We denote the common values
by $\cA$ and ${\cal C}^2$:
\be\label{cal_C} \cA = \cA_G = \cA_T, \qquad  {\cal C}^2 =
{\cal C}_G^2 = {\cal C}_T^2. \ee

The linearized Einstein field equations are given by
\be\label{lin_efe} {}^{(1)}\!G^a\!_b = {}^{(1)}\!T^a\!_b.   \ee
In simplifying the linearized field equations we will make use
of the intrinsic gauge invariants associated with the Einstein
tensor and with the stress-energy tensor, which are given, in
analogy with~\eqref{hatA}, \eqref{bf_A}
and~\eqref{bf_Ai}, by
\begin{subequations}\label{GT}
\begin{xalignat}{2}
\hat{\bf G}^i\!_j &=  a^2\,{}^{(1)}\!{\hat G}^i\!_j
&\quad  \hat{\bf T}^i\!_j &=  a^2\,{}^{(1)}\!{\hat T}^i\!_j \\
{\bf G}_i &= - a^2\left( {\bf D}_i {}^{(1)}\!G^0\!_0 + 3{\cal
H}{}^{(1)}\!G^0\!_i \right),
&\quad  {\bf T}_i &= -  a^2\left( {\bf D}_i {}^{(1)}\!T^0\!_0 +
3{\cal H}{}^{(1)}\!T^0\!_i \right), \label{GiTi}\\
{\bf G} &= a^2({\cal C}_G^2 {}^{(1)}\!G^0\!_0 + \sfrac13
{}^{(1)}\!G^k\!_k), &\quad {\bf T} &= a^2({\cal C}_T^2
{}^{(1)}\!T^0\!_0 + \sfrac13 {}^{(1)}\!T^k\!_k)  \label{bf_T1},
\end{xalignat}
\end{subequations}
where
\be  \label{GT_hat}  {}^{(1)}\!{\hat G}^i\!_j=
{}^{(1)}\!G^i\!_j - \sfrac13\delta^i\!_j {}^{(1)}\!G^k\!_k,
\qquad {}^{(1)}\!{\hat T}^i\!_j= {}^{(1)}\!T^i\!_j -
\sfrac13\delta^i\!_j {}^{(1)}\!T^k\!_k.  \ee
We also need the hybrid gauge invariants ${\bf G}^0\!_i  [X]$
and ${\bf T}^0\!_i  [X]$, which are given by~\eqref{AiX}
with $A$ replaced by $G$ and $T$:
\be \label{bf_G0i}  {\bf G}^0\!_i[X] = a^2\,{}^{(1)}G^0\!_i +
{\cal A}_G {\bf D}_i X^0, \qquad {\bf T}^0\!_i[X] =
a^2\,{}^{(1)}T^0\!_i + {\cal A}_T {\bf D}_i X^0. \ee

Since the gauge invariants~\eqref{GT} and~\eqref{bf_G0i} are
linear in $^{(1)}\!G^{a}\!_{b}$ and $^{(1)}\!T^{a}\!_{b}$ with
coefficients depending on ${}^{(0)}\!G^a\!_b$ and
${}^{(0)}\!T^a\!_b$, respectively, it follows that the
linearized Einstein field equations immediately imply the
following relations:
\begin{subequations}\label{linEFEtot}
\be {\bf\hat G}^{i}\!_{j} - {\bf \hat T}^{i}\!_{j} = 0, \qquad
{\bf G}_i - {\bf T}_i = 0,  \qquad {\bf G} - {\bf T} = 0,
\label{linEFE} \ee
\be \,\,\,\,\, {\bf G}^{0}\!_{i}[X] - {\bf T}^{0}\!_{i}[X] = 0.
\label{linEFE4} \ee
\end{subequations}
Expressions for the Einstein gauge invariants ${\hat{\bf
G}}{}^i\!_j, {\bf G}_i, {\bf G}$ and ${\bf G}^0\!_i[X] $ in
terms of the metric gauge invariants, decomposed into scalar,
vector, and tensor modes, are given in
equations~\eqref{intrinsiccurv} and~\eqref{G0i} in
Appendix~\ref{app:curv}. To proceed we likewise decompose the
matter gauge invariants ${\hat{\bf T}}{}^i\!_j, {\bf T}_i, {\bf
T}$ and ${\bf T}^0\!_i[X] $ into scalar, vector, and tensor
modes and label them as follows:\footnote{In
subsection~\ref{subsec:interpret} we comment on the choice of
the symbols $\Pi$, $\Gamma$, $\Delta$ and $V$.}
\begin{subequations}\label{T_i}       
\begin{align}
{\hat{\bf T}}{}^i\!_j &= {\bf D}^i\!_j \Pi +
2\gamma^{ik}{\bf D}_{(k}\Pi_{j)} + {\Pi}^i\!_j , \label{bf_T_hat}\\
{\bf T}_i & = {\bf D}_i \Delta+ \Delta_i, \label{bf_T_i} \\
{\bf T} & = \Gamma, \label{bf_T}\\
{\bf T}^0\!_i[X] & = {\bf D}_i V[X] + V_i, \label{hybrid_T}
\end{align}
where
\begin{equation}
{\bf D}^i\Pi_i =0 ,\qquad {\Pi}^k\!_k = 0 ,\qquad  {\bf
D}_i{\Pi}^i\!_j = 0, \qquad {\bf D}^i \Delta_i = 0, \qquad
{\bf D}^i V_i=0, \end{equation}
and
\be\label{Dij}    {\bf D}_{ij} := {\bf D}_{(i}{\bf D}_{j)} -
\sfrac13\gamma_{ij}{\bf D}^2, \qquad {\bf D}^2 := {\bf D}^i{\bf
D}_i. \ee
\end{subequations}
We stress that in making this decomposition we are not making
any assumptions about the physical nature of the stress-energy
tensor. By inspecting~\eqref{intrinsiccurv},~\eqref{G0i}
and~\eqref{T_i} one concludes that equations \eqref{linEFEtot}
decompose into a scalar mode, a vector mode and a tensor mode,
which we label as follows:
\begin{align*}
{\bf D}_{ij}\mathbb{A}
+ {\bf D}_{(i}\mathbb{A}_{j)} + \mathbb{A} _{ij} &= 0,\\
{\bf D}_{i}\mathbb{B} + \mathbb{B}_i &= 0,\\
\mathbb{C} &= 0,  \\
{\bf D}_{i}\mathbb{E}[X] + \mathbb{E}_i &= 0.
\end{align*}
Since we are assuming that the inverses of the operators ${\bf
D}^2, {\bf D}^2 + 2K$ and ${\bf D}^2 + 3K$ exist we can use the
proposition in Appendix~\ref{app:prop} to write the linearized
field equations concisely as
\begin{subequations}
\begin{align}
\text{{\it Scalar mode:}}\quad & \mathbb{A} \,\,= 0, \qquad\,\, \mathbb{B}
= 0, \qquad\, \mathbb{C} = 0, \qquad \mathbb{E}[X] = 0. \label{scalar_efe}\\
\text{{\it Vector mode:}}\quad & \mathbb{A}_i \, = 0,
\qquad \mathbb{B}_i = 0, \qquad \mathbb{E}_i = 0. \label{vector_efe}\\
\text{{\it Tensor mode:}}\quad & \mathbb{A}_{ij} = 0.
\label{tensor_efe}
\end{align}
\end{subequations}

\subsection{Scalar mode}

In this subsection we give the governing
equations~\eqref{scalar_efe} for the scalar mode, first
expressing them in terms of the uniform curvature gauge
invariants ${\bf A} = \Phi[X_\mathrm{c}]$ and ${\bf B} = {\bf
B}[X_\mathrm{c}]$ (see~\eqref{curv_scalar}). The scalars
$\mathbb{A}, \mathbb{B}$ and $ \mathbb{C}$ in
\eqref{scalar_efe} are obtained without any calculation by
taking the differences of equations~\eqref{intrinsiccurv} and
\eqref{T_i} and reading off the scalar part. The scalar
$\mathbb{E}[X]$ is obtained in a similar manner
from~\eqref{G0i} and~\eqref{hybrid_T} with $X= X_\mathrm{p}$.
The resulting equations are\footnote{In deriving
\eqref{Phicurv_evol} we use \eqref{bfB_evol} to replace
$\left(\partial_\eta + 2{\cal H}\right){\bf B} + {\bf A}$ by $
- \Pi$.}
\begin{subequations} \label{scalar_eq_curv}
\begin{align}
\left(\partial_\eta + 2{\cal H}\right){\bf B} + {\bf A} &= - \Pi\,\label{bfB_evol}\\
\cH \left[(\partial_{\eta} + {\cal BH}){\bf A} + {\cal C}_G^2 {\bf D}^2{\bf B}\right]
&= \sfrac12\Gamma + \sfrac13{\bf D}^2\Pi, \label{Phicurv_evol}\\
{\cal H}\left({\bf D}^2 + 3K\right){\bf B} &= - \sfrac12\Delta, \label{Poisson_C} \\
\cH {\bf A} + (\sfrac12\cA_G - K){\bf B} &= - \sfrac12{V}, \label{V_F}
\end{align}
\end{subequations}
where
\be  \label{cB} \cB = \frac{2{\cH}'}{ {\cal H}^2} + 1 +
3\cC_G^2, \ee
(see equation~\eqref{script_B} in Appendix~\ref{app:curv}), and
${V}={V}[X_\mathrm{p}]$. We shall refer to these equations as
the {\it uniform curvature form} of the governing equations for
the scalar mode.

We now give the governing equations in terms for the Poisson
gauge invariants $\Psi$ and $\Phi$. We eliminate ${\bf A}$
in~\eqref{Phicurv_evol} using~\eqref{bfB_evol} and
in~\eqref{V_F} using~\eqref{curv_Poisson}, and eliminate ${\bf
B}$ using $\cH {\bf B} = - \Psi$. The resulting equations are
\begin{subequations}\label{scalar_eq_Poisson}
\begin{align}
\Psi - \Phi &= \Pi, \label{phi} \\
\left({\bf {\cal L}} - {\cal C}_G^2{\bf D}^2\right)\Psi
&= \sfrac12\Gamma + \left(\sfrac13{\bf D}^2  + {\cal H} (\partial_{\eta}
+ {\cal BH}) \right) \Pi, \label{bardeen} \\
({\bf D}^2 + 3K)\Psi &= \sfrac12 \Delta, \label{Poisson_P} \\
\partial_{\eta}\Psi+ {\cal H}\Phi &= - \sfrac12{V},  \label{V_P}
\end{align}
\end{subequations}
where the differential operator ${\bf{\cal L}}$ is defined by
\be\label{factorL_s} {\bf {\cal L}}(\bullet) := {\cal H}(
\partial_\eta + {\cal BH} )(\partial_\eta + 2{\cal
H})\left(\frac{\bullet}{{\cal H}} \right), \ee
and ${\cB}$ is given by~\eqref{cB}.
Expanding the brackets yields\footnote{Referring
to~\eqref{ATCT} to express $\cH'$ in terms of $\cA_G$ and then
use the equation for $\cA_G'$.}
\be\label{L_s}  {\bf {\cal L}} =
\partial_\eta^2 + 3\left(1 + {\cal C}_G^2\right){\cal
H}\partial_\eta +  {\cal H}^2{\cal B} -(1 + 3\cC_G^2)K . \ee
We shall refer to the above equations as the {\it Poisson
form} of the governing equations for the scalar mode,
and to the evolution equation~\eqref{bardeen} as the
{\it Bardeen equation}.

Equations~\eqref{scalar_eq_curv} and~\eqref{scalar_eq_Poisson},
which are linked by the factorization property~\eqref{factorL_s},
constitute one of the main results of this paper.
Either system of equations determine the behaviour
of linear scalar perturbations of an FL cosmology with
arbitrary stress-energy content whose scalar mode is described
by the gauge invariants $\Gamma,\Pi, \Delta$ and ${V}$.
The
structure of these two systems of equations differs in a
significant way. In the system~\eqref{scalar_eq_curv} the time
dependence is governed by two {\it first order} differential
operators $\partial_{\eta} + {\cal BH}$ and $\partial_{\eta} + 2{\cal H}$,
 while in the system~\eqref{scalar_eq_Poisson} the
time dependence is governed by the {\it second order} linear
differential operator ${\cal L}$. A key point is that
{\it the coefficients in these operators
 depend only on the background RW
 geometry}, and this dependence manifests itself through
 the appearance of $\cH, \cH', \cH''$ and $K$.  This
property is significant since it means that these
operators will have the same form irrespective of the nature of
the source in the FL background model, {\it e.g.} whether it is
a perfect fluid with $p=p(\rho)$, or a scalar field with potential
$V(\phi)$. What will differ, however, is the functional
dependence of $\cH(\eta)$, which is determined by solving
the Einstein equations in the background RW geometry,
and hence depends on the source.
Furthermore these differential operators will also appear in the
linearized field equations in any geometrical theory of gravity,
whose field equations depend in some way on the Einstein tensor.

To the best of our knowledge equations~\eqref{scalar_eq_curv}
have not been given in the literature, although if one performs
a harmonic decomposition one obtains a system of first order
ordinary differential equations closely related to that given
by Kodama and Sasaki (1984)
(see Chapter 2, equations (4.6a-d)). Likewise, the governing
equations in Poisson form~\eqref{scalar_eq_Poisson}
have not appeared in the literature
in the above fully general form. The use of the Poisson gauge
invariants was initiated by Bardeen (1980), and the evolution
equation~\eqref{bardeen} for $\Psi$ is now commonly used,
although it is written in a variety of different forms, as a
partial or ordinary differential equation with the coefficients usually
expressed in terms of the matter variables of the background FL model.
In contrast we have written the Bardeen equation in a
fully general form in terms of the
purely geometric differential operator ${\bf{\cal L}} $, which is
defined by the factorization property~\eqref{factorL_s}.
We can relate our form of the equation to the literature by expanding
${\bf{\cal L}} $ as in~\eqref{L_s} and expressing the
coefficients in terms of the matter variables. If the matter
content is a  barotropic perfect fluid
 and a cosmological constant and one imposes the
 background Einstein field equations then the geometric coefficients
 $\cC_G^2$ and $\cB$ can be written as
\be  \cC_G^2 = c_s^2, \quad  {\cal H}^2\cB = (c_s^2- w)\rho a^2  + (1 +c_s^2)\Lambda a^2
- (1 + 3c_s^2) K,   \label{cB_pf} \ee
using~\eqref{GT0},~\eqref{cal_C} and~\eqref{c_s}.
The form in the literature that is
closest to the purely geometric form~\eqref{L_s}
is that given by Mukhanov {\it et al}
(1992), equation (5.22), who replace $\cC_G^2$ by the matter
quantity $c_{s}^2$ as in~\eqref{cB_pf} but
retain $\cH$ and $\cH'$. Nakamura (2007)
gives the same expression (see his equation (5.30)).
 A more common form in the literature
 has $\cB$, in addition to $\cC_G^2$, expressed in
terms of the background matter variables as in~\eqref{cB_pf}.
The earliest occurrence of which we are aware is Harrison
(1967), equation (182), followed by Bardeen (1980), equation
(5.30), after making the appropriate changes of notation and
setting $\Lambda = 0$. See also Ellis, Hwang and Bruni (1989),
equation (31) and Hwang and Vishniac (1990), equation
(105).\footnote{In these two references, the evolution equation
in question arises in the $1+3$ gauge-invariant approach to
perturbations of FL, and the unknown is a vector quantity that
is related to the scalar $\Psi$.}

\subsection{Vector and tensor modes}

First, we give the governing equations~\eqref{vector_efe} for
the {\it vector mode}. The vectors ${\mathbb A}_i$ and
${\mathbb B}_i$ in~\eqref{vector_efe}  are obtained without any
calculation by taking the differences of
equations~\eqref{intrinsiccurv} and~\eqref{T_i} and reading off
the vector part. The vector ${\mathbb E}_i$ is obtained in a
similar manner from~\eqref{G0i} and~\eqref{hybrid_T}. The
resulting equations are
\begin{subequations}\label{vector_eq}
\begin{align}
(\partial_\eta + 2{\cal H}) {\bf B}_i &= - 2\Pi_i,  \label{vector_evol}\\
({\bf D}^2 + 2K){\bf B}_i &=  2{V}_i, \label{vector_V}
\end{align}
\end{subequations}
as well as the relation $\Delta_i = 3{\cal H}{V}_i$, which is
satisfied identically (see equation~\eqref{gi_delta}). If
$\Pi_i$ is specified and can be regarded as a source term, the
evolution equation~\eqref{vector_evol} is a first order linear
ordinary differential equation that determines ${\bf B}_i$,
which in turn determines ${V}_i$ by differentiation
using~\eqref{vector_V}.

Second, we give the governing equations~\eqref{tensor_efe} for
the {\it tensor mode}. The tensor ${\mathbb A}_{ij}$
in~\eqref{tensor_efe} is obtained without any calculation by
taking the differences of equations~\eqref{intrinsiccurv}
and~\eqref{T_i} and reading off the tensor part, leading to
\be \label{tensor_evol}  \left(\partial_\eta^2 + 2{\cal
H}\partial_\eta + 2K - {\bf D}^2\right){\bf C}_{ij} =
{\Pi}_{ij}. \ee
If $\Pi_{ij}$ is specified and can be regarded as source term,
this is a second order linear partial differential equation
that determines ${\bf C}_{ij}$.

\section{Interpretations and examples}\label{sec:interpret}

\subsection{Interpretation of the matter gauge invariants}\label{subsec:interpret}

In this section we give the physical interpretation of the
gauge invariants $\Pi, \Gamma, \Delta$ and $V[X]$ associated
with the scalar mode of the stress-energy tensor.

We begin with the decomposition of a stress-energy tensor with
respect to a unit timelike vector field $u^a$, which is given by
\be \label{T_ab}  T^a\!_b = (\rho + p)u^a u_b + p\delta ^a\!_b +
  (q^au_b + u^aq_b) + \pi^a\!_b,   \ee
where
\be u^a q_b = 0,  \qquad  \pi^a\!_a = 0,  \qquad u_a \pi^a\!_b
= 0. \ee
We choose $u^a$ to be the timelike eigenvector of $T^a\!_b$,
which implies $q^a = 0$, {\it i.e.} we are using the so-called energy
frame (see for example, Bruni {\it et al} (1992), page 37).

Assuming that the unperturbed stress-energy tensor
${}^{(0)}T^a\!_b$ has the isotropy and homogeneity properties
of the RW geometry, the expansion~\eqref{taylor} to linear
order for $\rho, p$, $u_a$ and $\pi^a\!_b$ has the
form:\footnote{The form of $u_0$ is determined by the
requirement that $u^a$ is a unit vector. Recall that $\varphi$
is one of the metric potentials in~\eqref{split_f}.}
\begin{subequations}\label{Tdecomp}
\begin{xalignat}{2}
\rho &= {}^{(0)}\!\rho + \epsilon\,{}^{(1)}\!\rho, &\quad
p &= {}^{(0)}\!p + \epsilon\,{}^{(1)}\!p, \\
\pi^0\!_0 &= 0 =\pi^0\!_i, &\quad \pi^i\!_j &= 0 +
\epsilon\,{}^{(1)}\!\pi^i\!_j,\\
u_0 &= - a(1 +\epsilon\,\varphi), &\quad  u_i &= a( 0 +
\epsilon\,v_i).
\end{xalignat}
\end{subequations}
Decomposing $v_i$ into a scalar and vector mode yields
\be  \label{v_i} v_i = {\bf D}_i v + \tilde{\bf v}_i, \qquad
{\bf D}^i \tilde{\bf v}_i = 0.   \ee
We use boldface in writing $\tilde{\bf v}_i$ in view of the
fact that this quantity is a dimensionless gauge invariant, as
can be verified by applying~\eqref{delta_A} to $u_a$.

For ease of comparison with other work, we note that the
expansion of $u^a = g^{ab}u_b$ to linear order, expressed in
terms of $v$, $\tilde{\bf v}^i$ and the linearly perturbed
metric, is given by
\be   \label{contra_u}  u^0 =  a^{-1}(1 - \epsilon\,\varphi),
\qquad u^i = a^{-1}\left[0 +\epsilon\left({\bf D}^i(v - B) +
(\tilde{\bf v}^i - B^i)\right)\right].   \ee
We digress briefly to mention that our expansion of the
four-velocity differs from the usual approach in the literature
in that we use the {\it covariant} vector $u_a$ to define the
perturbed three-velocity instead of the contravariant vector
$u^a$, since we find that this leads to a number of
simplifications.\footnote{The source of these simplifications is
the fact that $u_i$ is invariant under purely spatial
gauge transformations while $u^i$ is not.}
 For example, Malik and Wands (2009) (see
equation (4.4)) have
\begin{equation*}
u^i = a^{-1}[0 +\epsilon\left({\bf D}^iv_{MW} + {\tilde
v}_{MW}^i\right)], \end{equation*}
so that
\begin{equation*}
v_{MW} = v - B, \qquad {\tilde v}_{MW}^i = \tilde{\bf v}^i -
B^i. \end{equation*}

From~\eqref{T_ab} and~\eqref{Tdecomp}, and making use
of~\eqref{perturb}, we obtain the following expressions for the
components of the linear perturbation of the stress-energy
tensor:
\be \label{pert_T} {}^{(1)}\!T^0\!_0 = - {}^{(1)}\!\rho, \qquad
{}^{(1)}\!T^k\!_k = 3\,{}^{(1)}\!p, \qquad {}^{(1)}\!T^0\!_i =
({}^{(0)}\!\rho + {}^{(0)}\!p)v_i, \qquad  {}^{(1)}\!{\hat
T}^i\!_j = {}^{(1)}\!\pi^i\!_j .  \ee
It follows from~\eqref{GT},~\eqref{T_i} and~\eqref{pert_T},
in conjunction with \eqref{ATCT} and \eqref{cons_T},
that the matter gauge invariants are determined by
\begin{subequations}\label{Tv}
\begin{align}
a^2 {}^{(1)}\!\pi^i\!_j &= {\bf D}^i\!_j \Pi +
2\gamma^{ik}{\bf D}_{(k}\Pi_{j)} + {\Pi}^i\!_j, \label{bf_pi_hat} \\
\Gamma &= a^2(-
{\cal C}_T^2 {}^{(1)}\!\rho + {}^{(1)}\!p), \label{pf_gamma}\\
\Delta &= a^ 2\left( {}^{(1)}\!\rho +
({}^{(0)}\!\rho)' \,v\right), \label{pf_delta}\\
{V}[X] &= {\cal A}_T (v + X^0),  \qquad\qquad\qquad     
{V}_i = {\cal A}_T \, \tilde{\bf v}_i.\label{pf_V}                  
\end{align}
\end{subequations}
Before continuing we derive an additional relation. It follows
from~\eqref{bf_Ai} with $A$ replaced by $T$ that
\be  {\bf T}_i = -  {\bf D}_i {\bf T}^0\!_0[X] -
3{\cal H}{\bf T}^0\!_i[X] .    \ee
On substituting from~\eqref{bf_T_i} and \eqref{hybrid_T}
into this equation, we conclude that
\be\label{gi_delta}  \Delta =  - {\bf T}^0\!_0[X] - 3\cH V[X],
\qquad \Delta _i = -3\cH {V}_i .      \ee

We can now give the physical interpretation of the matter gauge
invariants. First, the gauge invariants $\Pi, \Pi_i$ and
$\Pi_{ij}$ represent the anisotropic stresses. The
interpretation of $\Gamma$ is given in the context of a perfect
fluid in the next section. Next, the gauge invariants ${V} =
{V}[X_\mathrm{p}]$ and ${V}_i$ play a role in determining the
shear and vorticity of $u_a$. The relevant formulae are given
in~\eqref{kin} in Appendix~\ref{app:kin}. In particular,
${V}[X_\mathrm{p}]$ determines the scalar mode of the shear
according to
\be {\bf D}^j\!_i \bsigma^i\!_j = \sfrac23\cA_T^{-1}{\bf D}^2
({\bf D}^2 + 3K){V}[X_\mathrm{p}], \ee                                      
as follows from~\eqref{kin} in conjunction with~\eqref{pf_V}
with $X= X_\mathrm{p}$ and the identity~\eqref{i5}. We will
hence use ${V} := {V}[X_\mathrm{p}]$ as our standard choice for
the gauge invariant ${V}[X]$. However, since the choice
${V}[X_\mathrm{c}]$ is also of interest we note that
\be \label{V_C} {V}[X_\mathrm{c}] - {V}[X_\mathrm{p}] =
\cA_T {\bf B},  \ee                                                                    
as follows from~\eqref{pf_V},~\eqref{X_Poisson} and~\eqref{curv_scalar}.

Finally, in order to interpret $\Delta$ we need to make a small
digression. For any scalar field $A$ with the property that
$a^nA$ is dimensionless we can define a dimensionless gauge
invariant ${\bf A}[X]$ according to\footnote{This is equation
 \eqref{bold_A} specialized to the case of a scalar field.}
\be \label{bf_scalarA} {\bf A}[X] = a^n\left({}^{(1)}\!A -
({}^{(0)}\!{A}\,)^\prime\,{X}^0 \right). \ee
For the matter density $\rho$ we denote the gauge invariant
by $\brho[X]$:
\be\label{brho} \brho[X] = a^2\left({}^{(1)}\!\rho -
({}^{(0)}\!\rho\,)^\prime\,{X}^0 \right). \ee
On choosing $X = X_v$ with $X_v^0 := -v$ it follows
from~\eqref{pf_delta} that $\Delta = \brho[X_v]$. By
comparing~\eqref{brho} with equation (3.13) in Bardeen
(1980),\footnote{One has to take into account differences in
notation, the conservation equation~\eqref{cons_T}, and the
fact that Bardeen has performed a harmonic decomposition.} we
conclude that $\brho[X_v]$, and hence $\Delta$, equals the
well-known Bardeen gauge-invariant density perturbation
$\epsilon_m$, up to a factor of $a^2\,{}^{(0)}\!\rho$. The
specific relation is
\be \Delta = (a^2\,{}^{(0)}\!\rho) \epsilon_m.  \ee
We note that the choice $X_v^0 = -v$, in conjunction with our
default choice~\eqref{X_i} for the spatial components of $X$,
is associated with the so-called {\it total matter gauge} (see,
for example, Malik and Wands (2009), pages 23-24). Thus
$\Delta$ is the density perturbation in the total matter gauge.
In addition it turns out that $\Delta$ is closely related to
the $1+3$ gauge-invariant approach to perturbations of FL,
pioneered by Ellis and collaborators (see for example, Ellis and Bruni
(1989), Ellis {\it et al} (1989)), in which the spatial
gradient of the matter density orthogonal to $u^a$ plays a key
role. To elucidate the relation we define the dimensionless
spatial density gradient\footnote{Our ${\cal D}_a$ differs from
that in Bruni, Dunsby and Ellis (1992) by a factor of $\rho
a^2$ (see their equation (24)).}
\be {\cal D}_a(\epsilon) = a^2 h_a\!^b(\epsilon)\,
{}^\epsilon\bna\!_b \,\rho(\epsilon), \qquad h_a\!^b(\epsilon)
= \delta_a\!^b + u_a(\epsilon) u^b(\epsilon). \ee
A straight-forward calculation shows that ${\cal D}_a(0)=0$ and
that to linear order
\be  {}^{(1)}\!{\cal D}_0 = 0,  \qquad  {}^{(1)}\!{\cal D}_i =
{\bf D}_i \Delta - 3\cH {V}_i,  \ee
from which we conclude that $\Delta$ equals the the scalar mode
of the linear perturbation of the spatial density
gradient.\footnote{Note that ${}^\epsilon\bna\!_a
\,\rho(\epsilon)= {}^0\!\bar{\bna}\!_a \,\rho(\epsilon)$.} In
addition it follows from~\eqref{bf_T_i} and~\eqref{gi_delta}
that ${}^{(1)}\!{\cal D}_i = {\bf T}_i $, giving a physical
interpretation of the intrinsic gauge-invariant ${\bf T}_i$.

To end this section we comment on our choice of notation. In
using the symbols $\Pi, \Gamma, \Delta$ and $V$ for the matter
gauge invariants we are following Kodama and Sasaki (1984) with
the difference that we scale the variables as follows:
\be \Pi = a^2p\Pi_{KS}, \qquad \Gamma = a^2p\Gamma_{KS}, \qquad
\Delta = a^2\rho\Delta_{KS},  \qquad V = \cA_T V_{KS},  \ee                      
where $p$ and $\rho$ refer to the background. Our choice of
scalings simplify the equations considerably.

\subsection{Perfect fluid}

For a perfect fluid the matter gauge invariants are restricted
according to
\be\label{pf_pi} \Pi = 0, \qquad   \Pi_i = 0, \qquad  \Pi^i\!_j
= 0. \ee
In addition it follows from~\eqref{CT} and~\eqref{pf_gamma} that
\be \Gamma = 0 \quad \text{if and only if} \quad p=p(\rho), \ee
{\it i.e.} if and only if the equation of state is barotropic.
In this case it is customary to introduce the notation
\be \label{c_s}  c_s^2 := {\cal C}_T^2, \qquad
w :=\frac{^{(0)}\!p}{^{(0)}\!\rho}, \ee
where $c_s^2 = w$ if $w$ is constant, as follows from \eqref{CT}.

On account of~\eqref{pf_pi} the governing equations in the
Poisson form~\eqref{scalar_eq_Poisson} for scalar perturbations
imply that $\Psi - \Phi = 0$, which (in conjunction with the
background field equations) reduces the governing equations for
the scalar mode in the perfect fluid case to
\begin{subequations}\label{scalar_eq_Poisson_pf}
\begin{align}
({\bf {\cal L}} - c_s^2{\bf D}^2)\Psi &= \sfrac12 \Gamma, \label{bardeen_pf} \\
({\bf D}^2 + 3K)\Psi &= \sfrac12 \Delta, \label{Poisson_pf}  \\
\Psi^\prime + {\cal H}\Psi &= -\sfrac12{V}, \label{velocity_pf}
\end{align}
\end{subequations}
where ${\bf {\cal L}}$ is given by~\eqref{L_s} with ${\cal
C}_G^2={\cal C}_T^2=c_s^2$ and $\cB$ is expressed in terms of
the background matter variables according to~\eqref{cB_pf}.

\subsection{Scalar field}

For a minimally coupled scalar field we show in
Appendix~\ref{app:scalarfield} that the matter gauge invariants
are given by
\begin{subequations}\label{V,pi_scalar}
\begin{alignat}{2}
\Gamma &=  (1 - {\cal C}_T^2) \Delta, && \label{gamma_delta} \\
{V}[X] &= -{}^{(0)}\!\phi^\prime\bphi[X],
&\qquad\qquad  {V}_i &= 0, \label{V_scalar} \\
\Pi &= 0,  &\qquad\qquad  \Pi_i &= 0,  \qquad\qquad \Pi^i\!_j =
0, \label{pi_scalar}
\end{alignat}
\end{subequations}
where $\bphi[X]$ is the gauge invariant associated with
${}^{(1)}\!\phi$ by X-replacement, given by\footnote{This is a
special case of equation~\eqref{bf_scalarA}.}
\be \bphi[X] = {}^{(1)}\!\phi - {}^{(0)}\!\phi^\prime\, X^0. \ee
Note that $\cA_T$ and $\cC_T^2$ are given by \eqref{scriptA,C}.
The governing equations~\eqref{scalar_eq_Poisson} in
Poisson form imply that $\Psi - \Phi = 0$, and then reduce to
\begin{subequations}\label{scalareineqscalarfield}
\begin{align}
& ({\bf {\cal L}} - {\cal C}^2{\bf D}^2)\Psi
= \sfrac12 (1-{\cal C}^2)\Delta, \label{scalarfieldevol1}\\
& ({\bf D}^2 + 3K)\Psi = \sfrac12\Delta, \label{scalarfield3}\\
& \Psi^\prime + {\cal H}\Psi
= \sfrac12 {}^{(0)}\!\phi^\prime \bphi_\mathrm{p}, \label{einscalarV}
\end{align}
\end{subequations}
where $\bphi_\mathrm{p}:= \bphi[X_\mathrm{p}]$, and where we
have used $\cC^2_G = \cC^2_T = \cC^2$. By
combining~\eqref{scalarfieldevol1} and~\eqref{scalarfield3} we
obtain an evolution equation for $\Psi$ without a source term:
\be \left({\bf {\cal L}} - 3(1-{\cal C}^2)K - {\bf D}^2
\right)\Psi = 0,  \label{scalarevolmaster} \ee
where ${\bf {\cal L}}$ is given by~\eqref{L_s}. Having solved
this equation one can calculate $\bphi_\mathrm{p}$ and $\Delta$
from~\eqref{scalareineqscalarfield}. If one expresses $\cC^2$
in  ${\bf {\cal L}}$ in terms of the unperturbed scalar field
and its derivatives (see~\eqref{scriptA,C}) and sets $K=0$,
equation \eqref{scalarevolmaster} coincides with equation
(6.48) in Mukhanov {\it et al} (1992). For the generalization
to arbitrary $K$, see Nakamura (2007), equation
(5.39).\footnote{We note a minor typo: a factor of $2$
multiplying $\partial_\eta^2$ should be deleted.}

One can also use the governing equations~\eqref{scalar_eq_curv}
in uniform curvature form, obtaining equations equivalent to
those derived by Malik (2007) (see equations (2.20)-(2.23),
noting that he is considering multiple scalar fields).

\section{Discussion}\label{sec:discussion}

We have given a systematic account of the gauge-invariant
quantities that are associated with a linearly perturbed RW
geometry and stress-energy tensor, emphasizing the role of
intrinsic dimensionless gauge invariants. First, we have shown
that there are two distinct choices of dimensionless intrinsic
gauge invariants for the perturbed metric, which are the gauge
invariants associated with the Poisson gauge and the uniform
curvature gauge, through the work of Bardeen (1980) and Kodama
and Sasaki (1984), respectively. Second, we have introduced
dimensionless intrinsic gauge invariants for the Einstein
tensor and the stress-energy tensor, which we used to derive a
particularly simple and concise form of the governing equations
for linear perturbations of FL models. The specific form of the
governing equations for the scalar mode depends on the choice
of intrinsic gauge invariants for the perturbed metric. The
Kodama-Sasaki choice leads to a coupled system of two {\it
first order} (in time) linear differential operators that
govern the evolution of the uniform curvature metric gauge
invariants (see equations~\eqref{scalar_eq_curv}). On going
over to the Poisson picture, the product of these two operators
yields the {\it second order} linear differential operator
${\bf {\cal L}}$ that governs the evolution of the Bardeen
potential (see equation~\eqref{L_s}), thereby providing a link
between the two forms of the governing equations.
A common feature of both systems is the appearance of the
physically motivated gauge-invariant density perturbation
$\Delta$ that is one of the intrinsic gauge invariants
associated with the stress-energy tensor (see
equations~\eqref{Poisson_C} and~\eqref{Poisson_P}).

The mathematical structure of the governing equations for
linear perturbations that we have elucidated here has in fact a
much wider significance. Indeed, as one might expect on the
basis of elementary perturbation theory, the governing
equations for second order (nonlinear) perturbations have
precisely the same form, apart from the inclusion of a source
term that depends quadratically on the linear metric
perturbation.\footnote {This behaviour has been noted in
general terms by Nakamura (2006), equations (38)-(39).} As an
illustration of this we give the form of the equations that
govern second order scalar perturbations using the metric gauge
invariants associated with the Poisson gauge:
\begin{subequations} \label{scalar_eq_secondorder}
\begin{align}
{}^{(2)}\!\Psi - {}^{(2)}\!\Phi &= {}^{(2)}\!\Pi + S_{aniso} ({}^{(1)}\!{\bf f} ),  \\
\left({\bf {\cal L}} - {\cal C}_G^2{\bf D}^2\right){}^{(2)}\!\Psi
&= \sfrac12{}^{(2)}\Gamma + \left(\sfrac13{\bf D}^2  + {\cal H} (\partial_{\eta}
+ {\cal BH}) \right)\! {}^{(2)}\!\Pi + S_{evol} ({}^{(1)}\!{\bf f} ),  \label{psi_evol}   \\
({\bf D}^2 + 3K){}^{(2)}\!\Psi &= \sfrac12 {}^{(2)}\!\Delta +
S_{matter} ({}^{(1)}\!{\bf f} ), \label{matter}\\
\partial_{\eta}{}^{(2)}\!\Psi + {\cal H}{}^{(2)}\!\Phi &=
- \sfrac12{}^{(2)}\!V +  S_{velocity} ({}^{(1)}\!{\bf f} ),
\end{align}
\end{subequations}
where $S_{\bullet} ({}^{(1)}\!{\bf f} ) $ is a {\it source
term} that depends quadratically on the first order
gauge-invariant metric perturbation ${}^{(1)}\!{\bf f}_{ab}
\equiv {\bf f}_{ab}$ in equation~\eqref{bf_Poisson}. The key
point is that, apart from the source terms,
equations~\eqref{scalar_eq_secondorder} have the same form as
equations~\eqref{scalar_eq_Poisson}, with the variables
${}^{(2)}\!\Psi$ and ${}^{(2)}\!\Phi$ being the metric gauge
invariants at second order determined by the Nakamura
procedure. The second order matter terms ${}^{(2)}\!\Pi$,
${}^{(2)}\!\Gamma$, ${}^{(2)}\!\Delta$ and ${}^{(2)}\!V$ are
defined in analogy with the first order terms $\Pi$, $\Gamma$,
$\Delta$ and $V$ after expanding the stress-energy tensor
$T^a\!_b$ to second order in powers of $\epsilon$. All the
complications lie in the source terms, whose explicit form has
to be found by calculating the Riemann tensor to second order.
In order to solve the above second order equations the source
terms, which include scalar, vector and tensor modes, first
have to be obtained by solving the governing equations for the
scalar, vector and tensor linear perturbations. In a subsequent
paper we will derive both the above Poisson form and the
corresponding uniform curvature form of the governing equations
for second order perturbations, relating our formulation to
other recent work.

In this paper we have focussed exclusively on using the
linearized Einstein field equations to describe the dynamics of
scalar perturbations. There are, however, two alternatives to
the direct use of the linearized Einstein equations. First, one
can use the linearized conservation equations for the
stress-energy tensor, and second, one can use the $1+3$
gauge-invariant formalism,\footnote{See Bruni {\it et al}
(1992) for a comprehensive treatment.} in which the evolution
equations are obtained from the Ricci identities. An advantage
of using the first approach independently of the Einstein
equations is that the results are applicable to theories of
gravity other than general relativity. An advantage of the
second approach is that one initially derives exact nonlinear
evolution equations, which are then subsequently linearized.
Both of these approaches lead to a system of first order
partial differential equations that describe the evolution of
scalar perturbations. An additional aspect of the dynamics of
scalar perturbations that we have likewise not touched on in
this paper is that under certain conditions
 ({\it i.e.} in the long wavelength regime)
the governing equations admit so-called conserved quantities, {\it i.e.}
quantities that remain approximately constant during
a restricted epoch. These quantities, which are related to
both the linearized Einstein equations and the
linearized conservation equations, have been found
to be useful in analyzing the dynamics of scalar
perturbations during inflation. We refer to Uggla and
Wainwright (2011), where we discuss the above
aspects of the dynamics of scalar perturbations within
the framework of the present paper.

\subsection*{Acknowledgments}
CU is supported by the Swedish Research Council. CU also thanks
the Department of Applied Mathematics at the University of
Waterloo for kind hospitality. JW acknowledges financial
support from the University of Waterloo. We thank Henk van Elst
for helpful comments on a draft of this paper. We also thank an
anonymous referee for a constructive detailed report.

\appendix

\section{The Replacement Principle}\label{app:repl}

The expression for the perturbation of the Riemann tensor given
in equation~\eqref{firstcurvmaster} in Appendix~\ref{app:curv},
can be written symbolically in the form:
\be a^2{}^{(1)}\!R^{ab}\!_{cd} =  {\mathsf L}^{ab}\!_{cd}(f),
\ee
where ${\mathsf L}^{ab}\!_{cd}$ is a linear operator and $f$ is
shorthand for $f_{ab}$. The Replacement Principle for the
Riemann curvature states that the gauge invariants associated
with ${}^{(1)}\!R^{ab}\!_{cd}$ and with $f_{ab}$ by
$X$-compensation are related by {\it the same} linear operator:
\be {\bf R}^{ab}\!_{cd}[X] =  {\mathsf L}^{ab}\!_{cd}({\bf
f}[X]), \ee
where ${\bf f}[X]$ is shorthand for ${\bf f}_{ab}[X]$.

This result is adapted from more general results given by
Nakamura (2005) (see in particular, his equations (3.12),
(3.15) and (3.23)). Similar results hold for the Einstein and
Weyl tensors. Use of the Replacement Principle in Appendix
\ref{app:curv} makes the transition from gauge-variant to
gauge-invariant equations particularly easy and transparent.

\section{Derivation of the curvature formula}\label{app:curv}

In this appendix we derive expressions for the Einstein gauge
invariants, namely, the three intrinsic gauge invariants
$\hat{\bf G}^i\!_j, {\bf G}_i$ and ${\bf G}$, and the single
hybrid gauge invariant $^{(1)}\!{ \bf G}^0\!_i[X]$, defined by
equations \eqref{GT} and \eqref{bf_G0i}. Our strategy
incorporates the following ideas:
\begin{itemize}
\item[i)] {\it Conformal structure.} We adapt to the
    conformal structure of the background geometry,
    determined by the scale factor $a$ of the RW metric,
    from the outset. In particular we create dimensionless
    quantities by multiplying with appropriate powers of
    $a$, which simplifies the equations considerably.
 \item[ii)] {\it Index conventions.} We represent tensors
     of even rank, apart from the metric tensor, with equal
     numbers of covariant and contravariant indices. This
     makes contractions trivial to perform and ensures that
     the components of the tensor have the same physical
     dimension as the associated contracted scalar.
  \item[iii)] {\it Timing of specialization.} We defer
    performing the decomposition into scalar, vector and
    tensor modes as long as possible, and
    do not make harmonic decompositions.
     This strategy helps to
    reveal structure in the equations and serves to reduce
    the amount of calculation.
\end{itemize}
\subsubsection*{Calculation of $R^{ab}\!_{cd}(\epsilon)$}

We begin by deriving an exact expression for the Riemann
tensor\footnote{We use the sign convention of Wald (1984) for
defining the Riemann tensor.} $R^{ab}\!_{cd}(\epsilon)$  of the
metric $g_{ab}(\epsilon)$ in terms of the covariant derivative
of the conformal background metric $\gamma_{ab}$. We thus
relate the covariant derivative of $g_{ab}(\epsilon)$, denoted
${}^\epsilon\bna\!_a$, to that of
$\gamma_{ab}=\bar{g}_{ab}(0)$, denoted ${}^0\!\bar{\bna}\!_a$.
The relation is given by an object $Q^a\!_{bc} = Q^a\!_{cb} $
defined by
\be\label{metricQ} Q^a\!_{bc} = g^{ad}Q_{dbc} =
\sfrac{1}{2}g^{ad}\left({}^0\!\bar{\bna}\!_c g_{db} -
{}^0\!\bar{\bna}\!_d g_{bc} + {}^0\!\bar{\bna}\!_b
g_{cd}\right), \ee
(see Wald (1984) equation (D.1)), with the property
that\footnote{This example establishes the pattern for a
general tensor.}
\be \label{def_Q} {}^\epsilon\bna\!_a A^{b}\!_{c} =
{}^0\!\bar{\bna}\!_a A^{b}\!_{c} + Q^{b}\!_{ad}A^{d}\!_{c} -
Q^{d}\!_{ac}A^{b}\!_{d}.   \ee
It is convenient to write $Q^a\!_{bc}$ as the sum of
two parts:
\be \label{Qcosmo}  Q^a\!_{bc}(\epsilon) = \bar{Q}^a\!_{bc}(\epsilon) +
\tilde{Q}^a\!_{bc}(\epsilon). \ee
First, the transformation from $^\epsilon\bna\!_a$ to
$^\epsilon\bar{\bna}\!_a$, which is associated with the
conformal transformation $g_{ab}(\epsilon) =
a^2\bar{g}_{ab}(\epsilon)$, is described by
\be \bar{Q}^a\!_{bc} (\epsilon) = 2\delta^a\!_{(b} r_{c)} -
\bar{g}^{ad}(\epsilon)\bar{g}_{bc}(\epsilon) r_d, \ee
where\footnote{Note that we always use the vector $r_a$ in
covariant form, since $r_a$ is independent of $\epsilon$,
whereas $r^a = g^{ab}(\epsilon)r_b$ is not.}
\be \qquad r_a: = {}^0\!\bar{\bna}\!_a (\ln a) \ee
(see Wald (1984), equation (D.3)). It follows that
${}^0\!\bar{\bna}\!_a r_b = {}^0\!\bar{\bna}\!_b r_a.$ Second,
the transformation from $\!^\epsilon\bar{\bna}\!_a$ to
${}^0\!\bar{\bna}\!_a$, the covariant derivatives associated
with $\bar{g}_{ab}(\epsilon)$ and $\bar{g}_{ab}(0)$,
respectively, is described by
\be  \label{Qtilde} \tilde{Q}^a\!_{bc} (\epsilon) =
\sfrac{1}{2}\,\bar{g}^{ad}(\epsilon)\left({}^0\!\bar{\bna}\!_{c}\,\bar{g}_{db}(\epsilon)
- {}^0\!\bar{\bna}\!_{d}\,\bar{g}_{bc}(\epsilon) +
{}^0\!\bar{\bna}\!_{b}\,\bar{g}_{cd}(\epsilon)\right). \ee
It follows from ${}^0\!\bar{\bna}\!_a \gamma_{bc}=0$ that
\be \label{Qtilde0} \tilde{Q}^a\!_{bc}(0) =0. \ee

To calculate $R^{ab}\!_{cd}(\epsilon)$ we first perform the
conformal transformation from $g_{ab}$ to $\bar{g}_{ab}$, which
yields
\be\label{RiemQ2} a^2R^{ab}\!_{cd}(\epsilon) =
\bar{R}^{ab}\!_{cd}(\epsilon) + 4\delta^{[a}\!_{[c}
\bar{U}^{b]}\!_{d]}(\epsilon), \ee
where
\be \bar{U}^{b}\!_{d}(\epsilon) = - \left[\bar{g}^{be}\,(
{}^{\epsilon}\bar{\bna}\!_{d} - r_{d}) +
\sfrac{1}{2}\delta^{b}\!_{d}\,\bar{g}^{ef}\,r_f \right]r_e,
\label{Ueps1} \ee
and $\bar{R}^{ab}\!_{cd}(\epsilon)$ is the curvature tensor of
the metric $\bar{g}_{ab}(\epsilon)$ (see Wald (1984), equation
(D.7)). Second, by performing the transition from
$\!^\epsilon\bar{\bna}\!_a$ to ${}^0\!\bar{\bna}\!_a$ we obtain
\be \bar{R}^{ab}\!_{cd}(\epsilon) =
\bar{g}^{be}\bar{R}^{a}\!_{ecd}(\epsilon) =
\bar{g}^{be}\left({}^0\!\bar{R}^{a}\!_{ecd} +
2{}^0\!\bar{\bna}\!_{[c }\tilde{Q}^a\!_{d]e} +
2\tilde{Q}^a\!_{f[c}\tilde{Q}^f\!_{d]e}\right), \label{Riemeps}
\ee
where ${}^0\!\bar{R}^a{}_{bcd}$ is the curvature tensor of the
metric $\gamma_{ab}$ (see Wald (1984), equation (D.7)). The
term
$2\bar{g}^{be}\,{}^0\!\bar{\bna}\!_{[c}\tilde{Q}^a\!_{d]e}$
in~\eqref{Riemeps} can be written as\footnote{Note that
${}^0\!\bar{R}^{ab}{}_{cd} =
\gamma^{be}\,{}^0\!\bar{R}^{a}{}_{ecd}.$}
\begin{align}
\begin{split}
2\bar{g}^{be}\,{}^0\!\bar{\bna}\!_{[c }\tilde{Q}^a\!_{d]e} &=
2\bar{g}^{be}\left({}^0\!\bar{\bna}\!_{[c }\,\bar{g}^{af}\right)
\tilde{Q}_{|f|d]e} \\
& \quad +\, \bar{g}^{be}\,\bar{g}^{af}
({}^0\!\bar{\bna}\!_{[c}{}^0\!\bar{\bna}\!_{|e|}\,\bar{g}_{d]f} -
{}^0\!\bar{\bna}\!_{[c}{}^0\!\bar{\bna}\!_{|f|}\,\bar{g}_{d]e})
- \gamma_{ef} \bar{g}^{e(b}\,{}^0\!\bar{R}^{a)f}\!_{cd},
\end{split}
\end{align}
which we use to rearrange~\eqref{Riemeps}, 
in conjunction with  the relation
${}^0\!\bar{\bna}\!_c\bar{g}^{ab} = -2\tilde{Q}^{(ab)}\!_c$.
In summary, $R^{ab}\!_{cd}(\epsilon)$ is given by
equation~\eqref{RiemQ2} with
\begin{subequations}\label{mastercurv}
\begin{align}
\bar{R}^{ab}\!_{cd}(\epsilon) &=
-2\bar{g}^{e[a}\bar{g}^{b]f}
\,{}^0\!\bar{\bna}\!_{[c}{}^0\!\bar{\bna}\!_{|e|}\,\bar{g}_{d]f} -
\gamma_{ef}\bar{g}^{e[a}\,{}^0\!\bar{R}^{b]f}\,_{cd} -
2\tilde{Q}^{f[a}\!_{[c}\,\tilde{Q}_{|f|}\!^{b]}\!_{d]},
\label{Rieeps}\\
\bar{U}^{b}\!_{d}(\epsilon) &= -
\left[\bar{g}^{be}\,( {}^0\!\bar{\bna}\!_{d} - r_{d}) +
\sfrac{1}{2}\delta^{b}\!_{d}\,\bar{g}^{ef}\,r_f -
\bar{g}^{bf}\,\tilde{Q}^e\!_{df} \right]r_e, \label{Ueps}
\end{align}
\end{subequations}
where we have used ${}^\epsilon\bar{\bna}\!_a r_b =
{}^0\!\bar{\bna}\!_a r_b - \tilde{Q}^c\!_{ab}r_c$ in
obtaining~\eqref{Ueps} from~\eqref{Ueps1}.

\subsubsection*{Calculation of ${}^{(1)}\!R^{ab}\!_{cd}$}

We now calculate the perturbation ${}^{(1)}\!R^{ab}\!_{cd}$ of
the Riemann tensor, defined via equation~\eqref{perturb},
expressing it in terms of the covariant derivative
${}^{0}\!\bar{\bna}\!_a$ associated with $\gamma_{ab}$ and the
metric perturbation $ f_{ab} = {}^{(1)}\!\bar{g}_{ab}$
(see~\eqref{gamma,f}). We note that
\be\label{f_contra} {}^{(1)}\!{\bar g}^{ab} = -f^{ab}, \ee
where the indices on $f^{ab}$ are raised using $\gamma^{ab}$.
It follows from~\eqref{perturb}, \eqref{Qtilde}~\eqref{RiemQ2}
and~\eqref{mastercurv}, in conjunction with~\eqref{Qtilde0}
and~\eqref{f_contra}, that\footnote{Note that
$R^{ab}\!_{cd}(\epsilon)$ depends on $\epsilon$ through
$\bar{g}_{ab}(\epsilon), \bar{g}^{ab}(\epsilon)$ and $
\tilde{Q}^c\!_{ab}(\epsilon)$.}
\begin{subequations}\label{rie_taylor}
\be
a^2 {}^{(1)}\!R^{ab}\!_{cd} = {}^{(1)}\!\bar{R}^{ab}\!_{cd}  +
4\delta^{[a}\!_{[c}\!^{(1)}\bar{U}^{b]}\!_{d]},
\ee
where
\begin{align}
{}^{(1)}\!\bar{R}^{ab}\!_{cd} &=
- 2\,{}^0\!\bar{\bna}\!_{[c}{}^0\!\bar{\bna}{}^{[a}\,\,f_{d]}\!^{b]}
+ f_e\!^{[a}\,{}^0\!\bar{R}^{b]e}\!_{cd},\\
{}^{(1)}\!\bar{U}^{a}\!_{b} &= \left[f^{ac}\,(
{}^0\!\bar{\bna}\!_{b} - r_{b}) +
\sfrac{1}{2}\delta^{a}\!_{b}\,f^{cd}\,r_d +
\gamma^{ad}\,\,{}^{(1)}\!\tilde{Q}^c\!_{bd}\right]r_c, \\
{}^{(1)}\!\tilde{Q}_{abc} &=
\sfrac{1}{2}\left({}^0\!\bar{\bna}\!_{c}\,f_{ab} -
{}^0\!\bar{\bna}\!_{a}\,f_{bc} +
{}^0\!\bar{\bna}\!_{b}\,f_{ca}\right).
\end{align}
\end{subequations}
Introducing local coordinates $x^\mu=(\eta,x^i)$ as in
section~\ref{sec:genform} leads to
\be\label{r}  r_\alpha= {\cal H}\,\delta^0\!_\alpha , \qquad
{}^0\!\bar{\bna}\!_0 \,  = \partial_\eta , \qquad
{}^0\!\bar{\bna}\!_i  = {\bf D}_i . \ee
In addition we note that the quantity
${}^0\!\bar{R}^a{}_{bcd}$, the curvature tensor of the metric
$\gamma_{ab}$, is zero if one index is temporal, while if all
indices are spatial
\be\label{threecurv} {}^0\!\bar{R}^{ij}\!_{km} =
2K\delta^{[i}\!_{[k} \delta^{j]}\!_{m]}\, ,
\ee
where the constant $K$ describes the curvature of the maximally
symmetric three-space. Equation~\eqref{rie_taylor}, in
conjunction with~\eqref{r} and~\eqref{threecurv}, yields the
following expressions:
\begin{subequations}\label{firstcurvmaster}
\begin{align}
a^2{}^{(1)}\!{R}^{0 j}\!_{0 m} &= \sfrac{1}{2}[ {\bf D}^j{\bf D}_m +
({\cal H}^\prime - {\cal H}^2) \delta^j\!_m ]f_{00} + (\partial_\eta + {\cal H})Y^j\!_m, \\
a^2{}^{(1)}\!{R}^{0 j}\!_{km} &= 2{\bf D}_{[k} Y^j\!_{m]}, \\
a^2{}^{(1)}\!{R}^{ij}\!_{km} &=   -2\left({\bf D}_{[k}{\bf
D}^{[i} + K\delta_{[k}\!^{[i}\right)f_{m]}\!^{j]} + 4{\cal H}\delta_{[k}\!^{[i} Y_{m]}\!^{j]},
\end{align}
where\footnote{Note that $\tilde{Q}^0\!_{ij} =  - {\bf
D}_{(i}f_{j)0} + \sfrac{1}{2}f_{ij}^\prime$.}
\be Y_{ij} = \sfrac{1}{2}\gamma_{ij}{\cal H}f_{00} - {\bf
D}_{(i}f_{j)0} + \sfrac{1}{2}\partial_\eta f_{ij} \label{Y_ij}.
\ee
\end{subequations}

\subsubsection*{Calculation of the Riemann gauge invariants}

We now apply the Replacement Principle
to~\eqref{firstcurvmaster}, which entails performing the
following replacements:
\be  f_{ab} \rightarrow {\bf f}_{ab}[X],  \qquad Y_{ij}
\rightarrow {\bf Y}_{ij}[X], \qquad
a^2{}^{(1)}\!{R}^{ab}\!_{cd} \rightarrow {\bf
R}^{ab}\!_{cd}[X], \ee
where the gauge invariants are defined by
equation~\eqref{bold_A}. All components of the Riemann tensor
can be obtained from the `curvature spanning set'
($R^{0i}\!_{0j}$, $R^{0i}\!_{jk}$, $R^{im}\!_{jm}$) or,
alternatively, their spatial traces and their trace-free parts:
\be \label{curvspan}
(R^{0m}\!_{0m},\,R^{0m}\!_{jm},\,R^{km}\!_{km}),\qquad
(\hat{R}^{0i}\!_{0j},\,\hat{R}^{0i}\!_{jk},\,\hat{R}^{im}\!_{jm}),
\ee
where
\begin{subequations}
\be \hat{R}^{0i}\!_{0j} = R^{0i}\!_{0j} - \sfrac13 \delta^i\!_j
R^{0m}\!_{0m}, \qquad \hat{R}^{im}\!_{jm} = R^{im}\!_{jm} -
\sfrac13\delta^i\!_j R^{km}\!_{km}, \ee
\be \hat{R}^{0i}\!_{jk} = R^{0i}\!_{jk} -
\delta^i\!_{[k}R^{0m}\!_{j]m}. \ee
\end{subequations}
Our motivation for choosing these particular components as the
spanning set is that the first set of terms in~\eqref{curvspan}
are invariant under spatial gauge transformations, while the
hatted quantities are fully gauge-invariant, as follows
from~\eqref{delta_A}.

We denote the  gauge invariants associated with the spanning
set~\eqref{curvspan} by
\be \label{rie_gi} ({\bf R}^{0m}\!_{0m}[X],\,{\bf
R}^{0m}\!_{jm}[X],\,{\bf R}^{km}\!_{km}[X]), \qquad  (\hat{{\bf
R}}^{0i}\!_{0j},\,\hat{{\bf R}}^{0i}\!_{jk},\,\hat{{\bf
R}}^{im}\!_{jm}),  \ee
and refer to them as the {\it Riemann gauge invariants.} As
indicated by the notation ({\it i.e.}~no dependence on the
gauge field $X$) the hatted quantities are intrinsic gauge
invariants. We now substitute the expressions\footnote{In using
these expressions we are making the choice for $X_i$ given in
equation~\eqref{X_i}. Choosing $X_i$ in this way simplifies the
calculation but not the final form of the Riemann gauge
invariants, since, as mentioned earlier, the spanning set is
invariant under spatial gauge transformations. } for ${\bf
f}_{ab}[X]$ given by~\eqref{bold_f_split} into the bold-face
version of~\eqref{firstcurvmaster}, and calculate the gauge
invariants~\eqref{rie_gi}. It is convenient to split ${\bf
Y}_{ij}$ into a trace and a trace-free part:
\be \hat{{\bf Y}}_{ij} = {\bf Y}_{ij} - \sfrac13\gamma_{ij}
{\bf Y}, \qquad  {\bf Y} = {\bf Y}^i\!_i, \ee
and to use the trace-free second derivative operator ${\bf
D}_{ij}$ defined in~\eqref{Dij}. We obtain\footnote{Use the
identities~\eqref{i3}, \eqref{i4} and \eqref{i8}.  }
\begin{subequations}\label{secondcurvmaster}
\begin{align}
{\bf R}^{0m}\!_{0m}[X] &= - \left[{\bf D}^2 + 3({\cal H}^\prime - {\cal H}^2) \right]\Phi[X]
+ \left(\partial_\eta + {\cal H}\right){\bf Y}[X], \\
\hat{{\bf R}}^{0 i}\!_{0 j} &= - {\bf D}^i\!_j \Phi[X] +
\left(\partial_\eta + {\cal H}\right)\hat{{\bf Y}}^i\!_j[X], \\
{\bf R}^{km}\!_{km}[X] &= 4\left[\left({\bf D}^2 +
3K\right)\Psi[X] + {\cal H}{\bf Y}[X]\right], \\
\hat{{\bf R}}^{im}\!_{jm} &= {\bf D}^i\!_j  \Psi[X] + {\cal
H}\hat{{\bf Y}}^i\!_j[X] - \left({\bf D}^2 - 2K\right){\bf
C}^i\!_j,  \\
{\bf R}^{0m}\!_{jm}[X] &= \sfrac23{\bf D}_j {\bf Y}[X] -
{\bf D}_m \hat{{\bf Y}}^m\!_j[X], \\
\hat{{\bf R}}^{0 i}\!_{jk} &= 2{\bf D}_{[j} \hat{{\bf
Y}}^i\!_{k]}[X] + {\bf D}_m\hat{{\bf
Y}}^m\!_{[j}[X]\delta^i\!_{k]},
\end{align}
where
\begin{align}
{\bf Y}[X] &= -3(\partial_\eta \Psi[X] +{\cal H}\Phi[X]) -
{\bf D}^2 {\bf B}[X], \\
\hat{{\bf Y}}_{ij}[X] &= -  {\bf
D}_{ij}  {\bf B}[X] - {\bf D}_{(i}{\bf B}_{j)} +
\partial_{\eta}{\bf C}_{ij}.
\end{align}
\end{subequations}
These equations constitute one of the main results of this
paper. They express the Riemann gauge invariants \eqref{rie_gi}
in terms of the metric gauge invariants~\eqref{bold_f_split}.
They depend only on the choice of the temporal gauge field
$X^0$, as can be seen from~\eqref{generic_scalar}.

\subsubsection*{Calculation of the Einstein gauge invariants}

The  Einstein tensor and the Weyl conformal curvature tensor
are defined in terms of the Riemann tensor according to
\begin{subequations} \label{einweyl_def}
\begin{align}
G^a\!_b &:= R^a\!_b -\sfrac{1}{2}\,\delta^a\!_b R, \qquad\quad \text{where}\qquad\quad
R^a\!_b := R^{ac}\!_{bc},\qquad R :=  R^a\!_a,\\
C^{ab}\!_{cd} &:= R^{ab}\!_{cd} -
2\,\delta^{[a}\!_{[c}\,R^{b]}\!_{d]} + \sfrac{1}{3}\,
\delta^{[a}\!_{[c}\,\delta^{b]}\!_{d]}\,R . \lb{weyl_defn}
\end{align}
\end{subequations}
The curvature spanning set~\eqref{curvspan} can be replaced
with the following spatially irreducible components of the
Einstein tensor and the Weyl tensor:\footnote{Note that
$C^{ij}\!_{km} = -4C^{0[i}\!_{0[k}\,\delta^{j]}\!_{m]}$ in an
orthonormal frame.}
\be\label{einweyl} (G^0\!_0,\,
G^m\!_m,\,G^0_i,\,\hat{G}^i\!_j), \qquad
(C^{0i}\!_{0j},\,C^{0i}\!_{jk}), \ee
where
\be \hat{G}^i\!_j := G^i\!_j - \sfrac13 \delta^i\!_ j G^m\!_m. \ee
It follows from \eqref{einweyl_def} that
\begin{subequations}\label{einweylspan}
\begin{xalignat}{2}
G^0\!_0 &= -\sfrac12R^{km}\!_{km}, &\quad G^m\!_m & = -\sfrac12(R^{km}\!_{km} + 4R^{0m}\!_{0m}),\\
G^0\!_i &= R^{0m}\!_{im}, &\quad \hat{G}^i\!_j &= \hat{R}^{0i}\!_{0j} + \hat{R}^{im}\!_{jm},\\
C^{0i}\!_{0j} &= \sfrac12(\hat{R}^{0i}\!_{0j} -
\hat{R}^{im}\!_{jm}), & \quad  C^{0i}\!_{jk} &=
\hat{R}^{0i}\!_{jk}. \label{weyl_span}
\end{xalignat}
\end{subequations}
The Einstein gauge invariants, as defined by
equations~\eqref{hatA}, \eqref{bf_A} and~\eqref{bf_Ai} with $A$
replaced by $G$, can be expressed in terms of the curvature
spanning set~\eqref{curvspan} by using the bold-face version
of~\eqref{einweylspan}. This yields
\begin{subequations}\label{ein_inv}
\begin{align}
\hat{\bf G}^i\!_j & := {\hat{\bf G}}^i\!_j[X]= \hat{{\bf R}}^{0i}\!_{0j} +
\hat{{\bf R}}^{im}\!_{jm}, \\
{\bf G}_i &:= -\left( {\bf D}_i {\bf
G}^0\!_0 [X] + 3 {\cal H}{\bf G}^0\!_i [X]\right)=  \sfrac12 {\bf D}_i{\bf R}^{km}\!_{km}[X] -
3{\cal H}{\bf R}^{0m}\!_{im}[X], \\
{\bf G} &:= {\cal C}_G^2 {\bf G}^0\!_0 [X] +
\sfrac13 {\bf G}^m\!_m [X] = - \sfrac16 \left((1+ 3{\cal C}_G^2){\bf R}^{km}\!_{km}[X] +
4{\bf R}^{0m}\!_{0m}[X]\right).
\end{align}
\end{subequations}

We find that it is simplest to express the Einstein gauge
invariants \eqref{ein_inv} in terms of the {\it uniform
curvature} metric gauge invariants ${\bf A}$ and ${\bf B}$
defined by \eqref{curv_scalar}. We accomplish this directly by
choosing $X = X_\mathrm{c}$ in~\eqref{secondcurvmaster}, and
noting that by \eqref{choices} we have $\Psi[ X_\mathrm{c}] =
0$. After simplifying using the identities \eqref{i5} and
\eqref{i6} we obtain\footnote{Here for convenience we use
$\hat{\bf G}_{ij} = \gamma_{ik} \hat{\bf G}^k\!_j.$}
\begin{subequations}\label{intrinsiccurv}
\begin{align}
\hat{\bf G}_{ij} &= {\bf D}_{ij}  \mathbb{G}
- {\bf D}_{(i} \left(\partial_\eta + 2{\cal H}\right) {\bf B}_{j)} +
\left( \partial_\eta^2 + 2{\cal H} \partial_\eta  + 2K - {\bf D}^2\right){\bf C}_{ij}, \\
{\bf G}_i &=  2 \cH{\bf D}_i({\bf D}^2 + 3K){\bf B} + \sfrac32\cH({\bf D}^2 + 2K){\bf B}_i, \\
{\bf G} &=   2{\cal H}[(\partial_{\eta} + {\cal BH}){\bf A} +
{\cal C}_G^2 {\bf D}^2{\bf B}] -\sfrac23{\bf D}^2  \mathbb{G},
\end{align}
\end{subequations}
where we have introduced the notation
\be\label{script_B} \mathbb{G} := - [ {\bf A} + (\partial_\eta
+ 2{\cal H}){\bf B}], \qquad \cB :=   \frac{2 {\cH}'}{ {\cal
H}^2} +  1 + 3\cC_{G}^2 . \ee

We also need
\be {\bf G}^0\!_j[X] = {\bf R}^{0m}\!_{jm}[X] . \ee
We choose $X = X_\mathrm{p}$ in this equation, and  using~\eqref{secondcurvmaster}
in conjunction with the identity~\eqref{i6} we obtain
\be {\bf G}^0\!_j[X_\mathrm{p}] = - 2{\bf D}_j(\partial_\eta
\Psi + \cH \Phi) + \sfrac12\left({\bf D}^2 + 2K\right){\bf B}_j
. \ee
We now use~\eqref{curv_Poisson} to express the right side of
this equation in terms of ${\bf A}$ and ${\bf B}$,
which yields
\be\label{G0i}  {\bf G}^0\!_i[X_\mathrm{p}] = -2{\bf
D}_i\left(\cH {\bf A} + (\sfrac12\cA_G - K){\bf B}\right) +
\sfrac12\left({\bf D}^2 + 2K\right){\bf B}_i . \ee

\subsubsection*{The Weyl tensor}

The perturbation of the Weyl tensor is automatically
gauge-invariant on account of the Stewart-Walker lemma
(Stewart and Walker (1974)) since the Weyl tensor is zero in the
background. We thus use bold-face notation for its components.
From~\eqref{weyl_span} we obtain
\be {\bf C}^{0i}\!_{0j} = a^2{}^{(1)}\!C^{0i}\!_{0j} =
\sfrac12(\hat{{\bf R}}^{0i}\!_{0j} - \hat{{\bf
R}}^{im}\!_{jm}), \qquad {\bf C}^{0i}\!_{jk} =
a^2{}^{(1)}\!C^{0i}\!_{jk} =\hat{{\bf R}}^{0i}\!_{jk}. \ee
The Weyl tensor has a simpler form if we use Poisson
gauge invariants and hence we choose $X= X_\mathrm{p}$
in~\eqref{secondcurvmaster}. Noting that ${\bf B}[X_\mathrm{p}]
=0$ leads to
\begin{subequations}
\begin{align}\label{weyl_curv} {\bf C}^{0i}\!_{0j} &=
-\sfrac12\left[{\bf D}^i\!_j (\Psi + \Phi) + \partial_\eta {\bf
B}^i\!_j - \left(\partial_\eta^2 + {\bf D}^2 -
2K\right){\bf C}^i\!_j\right] ,\\
{\bf C}^{0i}\!_{jk} &= -2{\bf D}_{[j}
\left( {\bf B}^i\!_{k]} - \partial_\eta {\bf C}^i\!_{k]}\right)
- {\bf D}_m {\bf B}^m\!_{[j}\delta^i\!_{k]}, \qquad\qquad\qquad
{\bf B}_{ij} := {\bf D}_{[i} {\bf B}_{j]}.
\end{align}
\end{subequations}

\subsection{Uniqueness of the decomposition into
modes}\label{app:prop}

{\bf Proposition}: If the inverses of the operators ${\bf D}^2,
{\bf D}^2 + 2K$ and ${\bf D}^2 + 3K$ exist, then the equation
\be\label{p1} B_i = {\bf D}_i B + \tilde{B}_i, \quad
\text{with} \quad {\bf D}^i \tilde{B}_i =0, \ee
determines $B$ and $\tilde{B}_i$ uniquely in terms of $B_i$,
and the equation
\be\label{p2}  C_{ij} = {\bf D}_{ij}C + {\bf
D}_{(i}C_{j)} + \tilde{C}_{ij}, \ee
with
\begin{equation*}  {\bf D}^i C_i =0, \qquad \tilde{C}_{ij} = \tilde{C}_{ji},
\qquad \tilde{C}^i\!_i = 0, \qquad
{\bf D}^i \tilde{C}_{ij} = 0, \end{equation*}
determines $C$, $C_i$ and $\tilde{C}_{ij}$  uniquely in terms
of $C_{ij}$. In particular, if $B_i = 0$ then $B=0$,
$\tilde{B}_i = 0$, and if $C_{ij} = 0$ then $C = 0$, $C_i = 0$,
$\tilde{C}_{ij} = 0$.

\begin{proof}
Apply ${\bf D}^i$ to~\eqref{p1} obtaining ${\bf D}^i B_i = {\bf
D}^2 B$. Using the inverse operator of ${\bf D}^2$ this
equation determines $B$, and then~\eqref{p1} determines
$\tilde{B}_i$ uniquely in terms of $B_i$. Next, apply ${\bf
D}^{ij}$ and ${\bf D}^i$ to~\eqref{p2}, obtaining
\be {\bf D}^{ij}C_{ij} = \sfrac23 {\bf D}^2({\bf D}^2 + 3K)C,
\qquad {\bf D}^i C_{ij} = \sfrac23 {\bf D}_j ({\bf D}^2 + 3K)C
+ ({\bf D}^2 + 2K)C_j .\ee
By using the inverse operators these equations, in conjunction
with~\eqref{p2}, successively determine $C, C_i$ and
$\tilde{C}_{ij}$ uniquely in terms of $C_{ij}$.
\end{proof}

\subsection{Identities}

In obtaining our results we found the following identities
useful:
\begin{subequations}
\begin{align}
{\bf D}_{[i}{\bf D}_{j]}A^k &= K\delta^k\!_{[i}A_{j]}, \\
{\bf D}_{[k}{\bf D}_{m]} A^{ij} &= 2K \delta_{[k}\!^{(i} A_{m]}\!^{j)}, \\
4({\bf D}_{[k}{\bf D}^{[i} + K\delta_{[k}\!^{[i})\delta_{m]}\!^{m]}A &= \left({\bf D}_k\!^i +
\sfrac43\left({\bf D}^2 + 3K\right)\delta_k\!^i\right)A, \label{i3} \\
4\left({\bf D}_{[k}{\bf D}^{[i} +
K\delta_{[k}\!^{[i}\right)C_{j]}\!^{j]} &= ({\bf D}^2 - 2K)C^i\!_k, \label{i4}\\
{\bf D}_j{\bf D}^j\!_i A &= \sfrac23\, {\bf D}_i({\bf D}^2 + 3K)A, \label{i5}\\
{\bf D}^i {\bf D}_{(i}A_{j)} &= \sfrac12({\bf D}^2 + 2K)A_j,  \label{i6}\\
{\bf D}_i {\bf D}^2 A^i &= ({\bf D}^2 + 2K) {\bf D}_i A^i, \\
\delta_{[i}\!^{[i} A_{m]}\!^{j]} &= \sfrac{1}{4}(A_m\!^j + \delta_m\!^j\,A), \label{i8}
\end{align}
\end{subequations}
where $A_{ij} = A_{ji}$, $C_{ij} = C_{ji}$, $C^i\!_i=0$
and ${\bf D}_i C^i\!_j=0$.

\subsection{Kinematic quantities}\label{app:kin}

The kinematic quantities associated with a timelike congruence $u^a$
are defined by the following decomposition into irreducible parts:
\be \bna\!_au_b = - u_a {\dot u}_b + H(g_{ab} + u_au_b) +
\sigma_{ab}
 + \omega_{ab}. \ee
A routine calculation starting with
equations~\eqref{Tdecomp}-\eqref{contra_u} and \eqref{def_Q}
applied to $u_a$ yields the following non-zero components:
\begin{subequations} \label{kin}
\begin{align}
a{}^{(1)}\!H &= \left[ \sfrac13 {\bf D}^2 (v - \chi) -
(\partial_{\eta} \psi + \cH \varphi) \right],\\
\dot{\bf u}_i&:={}^{(1)}\!\dot{u}_i = {\bf D}_i \left(\varphi +
(\partial_\eta + \cH)v \right) +(\partial_\eta + \cH)\tilde{\bf v}_i ,\\
\bsigma ^i\!_j &:= a {}^{(1)}\!\sigma^i\!_j = {\bf D}^i\!_j (v - \chi) +
\gamma^{ik}{\bf D}_{(k}\left(\tilde{\bf v}_{j)} - {\bf B}_{j)} \right) +
\partial_\eta {\bf C}^i\!_j ,\\
\bomega^i\!_j &:= a{}^{(1)}\!\omega^i\!_j =  \gamma^{ik}{\bf D}_{[k}\tilde{\bf v}_{j]} ,
\end{align}
\end{subequations}
with the bold-face quantities being gauge-invariant on account
of the Stewart-Walker lemma.
%

\section{Scalar field}\label{app:scalarfield}

A minimally coupled scalar field $\phi$ is described by a
stress-energy tensor of the form
\be\label{sfT_ab} T^a\!_b = \bna^a\phi\bna\!_b\phi -
\left[\sfrac12 \bna^c\phi\bna\!_c\phi +
U(\phi)\right]\delta^a\!_b , \ee
with the associated Klein-Gordon equation $\bna^c\bna\!_c\phi -
U\!_{,\phi} = 0$, where the potential $U(\phi)$ has to be
specified. This stress-energy tensor is of the
form~\eqref{T_ab} with
\be \label{T_ab,scalar} \rho + p = -  \bna^a\phi\bna\!_a\phi,
\qquad \rho - p = 2U(\phi), \qquad \pi_{ab} = 0. \ee
When evaluated on the RW background,
equation~\eqref{T_ab,scalar} leads to
\be \label{rho,p for phi} a^2({}^{(0)}\!\rho + {}^{(0)}\!p) =
({}^{(0)}\!\phi^\prime)^2, \qquad  {}^{(0)}\!\rho - {}^{(0)}\!p
= 2U({}^{(0)}\!\phi) . \ee
On using~\eqref{rho,p for phi} to calculate ${}^{(0)}\!\rho'$,
the conservation equation~\eqref{cons_T} leads to
\be \label{KG}  { {}^{(0)}\!\phi}'' + 2{\cal H} {}^{(0)}\!\phi^\prime + a^2U\!_{,\phi}\, = 0,
\ee
which is the Klein-Gordon equation in the RW background.
Further, by means of~\eqref{ATCT},~\eqref{cons_T},~\eqref{rho,p
for phi} and~\eqref{KG} we obtain
\be\label{scriptA,C}  \mathcal{A}_T = ({}^{(0)}\!\phi')^2,
\qquad \mathcal{C}_T^2 = 1 + \frac{2a^2U_{,\phi} }
{3\mathcal{H}  {}^{(0)}\!\phi'} = -\sfrac13\left(1 + \frac{2
{}^{(0)}\!\phi''}{\cH {}^{(0)}\!\phi'}\right). \ee

Viewing $T^a\!_b$ and $\phi$ as functions of the perturbation
parameter $\epsilon$, we can use~\eqref{sfT_ab}, in conjunction
with~\eqref{perturb}, to calculate ${}^{(1)}\!T^a\!_b$,
obtaining
\be {}^{(1)}\!\hat T^i\!_j = 0, \qquad
a^2\,{}^{(1)}\!T^{0}\!_{i} = - {}^{(0)}\!\phi^\prime\, {\bf
D}_i {}^{(1)}\!\phi, \qquad {}^{(1)}\!T^0\!_0 + \sfrac13
{}^{(1)}\!T^i\!_i = -2U_{,\phi} {}^{(1)}\!\phi . \label{4.11}
\ee
It follows using~\eqref{DeltaAtens} with $A$ replaced by $T$
and \eqref{scriptA,C}, that the matter gauge invariants assume
the form
\be \label{bf_T_scalar}  {\bf \hat T}^i\!_j = 0, \qquad {\bf
T}^0\!_i[X] = - {}^{(0)}\!\phi^\prime\, {\bf D}_i \bphi[X],
\qquad  {\bf T}^0\!_0[X] + \sfrac13 {\bf T}^i\!_i[X] = -2a^2
U_{,\phi}\bphi[X],   \ee
where $\bphi[X]$ is the gauge invariant associated with
${}^{(1)}\!\phi$ by X-replacement, given by
\be \bphi[X] = {}^{(1)}\!\phi - {}^{(0)}\!\phi^\prime\, X^0. \ee
Equations~\eqref{bf_T_scalar} and~\eqref{T_i} immediately lead
to the expressions for the matter gauge
invariants~\eqref{V_scalar} and~\eqref{pi_scalar}, including
\be V[X] = - {}^{(0)}\!\phi^\prime\,\bphi[X].  \label{V_temp}  \ee
Equation~\eqref{bf_T_scalar}, in conjunction with~\eqref{scriptA,C}
and~\eqref{V_temp},  yields
\be\label{rho+p} {\bf T}^0\!_0[X] + \sfrac13 {\bf T}^i\!_i[X] =
-3(1 - {\cal C}_T^2) {\cal H}V[X].\ee
We now substitute~\eqref{rho+p} into the expression for
$\Gamma$ given by~\eqref{bf_T1} and~\eqref{bf_T} to
obtain\footnote{Write the expression for $\Gamma$ in the form
$\Gamma = - (1 - {\cal C}_T^2){\bf T}^0\!_0 + ({\bf T}^0\!_0 +
\sfrac13 {\bf T}^i\!_i).$}
\be\label{gd_temp} \Gamma =  (1 - {\cal C}_T^2)(- {\bf
T}^0\!_0[X] - 3\cH V[X]). \ee
which on comparison with~\eqref{gi_delta} leads to
equation~\eqref{gamma_delta}.

\section*{References}

\noindent Bardeen, J. M. (1980) Gauge-invariant cosmological
perturbations, {\it Phys. Rev. D} {\bf 22}, 1882-1905.\\

\noindent Bertschinger, E. (1996) Cosmological dynamics, in
{\it Cosmology and large scale structure}, eds. R. Schaeffer, J. Silk, M. Spiro
and J. Zinn-Justin, Elsevier Scientific Publishing Company,
(arXiv:astro-ph/9503125v1). \\

\noindent Brandenberger, R., Kahn, R. and Press, W.H. (1983)
Cosmological perturbations in the early universe,
{\it Phys. Rev. D} {\bf28},1809 - 1821. \\

\noindent Bruni, M., Dunsby, P.K.S. and Ellis, G.F.R. (1992)
Cosmological perturbations and the meaning of gauge-invariant
variables, {\it Astrophysical J.} {\bf 395}, 34-53.\\


\noindent Bruni, M., Matarrese, S., Mollerach, S., and Sonego,
S. (1997) Perturbations of spacetime: gauge transformations and
gauge-invariance at second order and beyond,
{\it Class. Quant. Grav.} {\bf 14} 2585-2606.\\

\noindent Christopherson, A. J., Malik, K. A., Matravers, D. R. and Nakamura, K. (2011)
Comparing different formulations of nonlinear perturbation theory,
arXiv:1101.3525.  \\

\noindent Durrer, R. (1994) Gauge-invariant cosmological
perturbation theory, {\it Fundamental Cosm. Phys} {\bf 15}, 209-.
(arXiv:astro-ph/9311041)\\

\noindent Durrer, R. (2008) {\it The Cosmic Microwave Background},
Cambridge University Press. \\

\noindent Eardley, D. M., (1974), Self-similar spacetimes:
Geometry and dynamics, {\it Commun. math. Phys} {\bf 37},
287-309. \\

\noindent Ellis, G.F.R. and Bruni, M. (1989), Covariant and
gauge-invariant approach to cosmological density
fluctuations, {\it Phys. Rev. D} {\bf 40}, 1804�1818.\\

\noindent Ellis, G.F.R., Hwang, J. and Bruni, M. (1989),
Covariant and gauge-independent perfect fluid Robertson-Walker
perturbations, {\it Phys. Rev. D} {\bf 40}, 1819-1826.\\

\noindent Heinzle, J. M., R\"ohr, N. and Uggla C. (2003),
Dynamical systems approach to relativistic spherically
symmetric static perfect fluid models, {\it Class. Quant.
Grav.} {\bf 20}, 4567-4586. \\

\noindent Hwang, J. and Vishniac, E.T. (1990)
Analyzing cosmological perturbations using the covariant approach,
{\it Astrophysical Journal} {\bf 353}, 1-20.    \\

\noindent Kodama, H. and Sasaki, M. (1984) Cosmological
Perturbation Theory, {\it Prog. Theoret. Phys. Suppl. } {\bf 78},1-166.\\

\noindent Liddle, A. R. and Lyth, D. H. (2000) {\it Cosmological Inflation and
Large-Scale Structure}, Cambridge University Press.  \\

\noindent Lifshitz, E. (1946) On the Gravitational Stability of the Expanding
Universe, {\it J. Phys. (Moscow)} {\bf 10}, 116-129. \\


\noindent Lyth, D. H. and Liddle, A. R. (2009) {\it
The Primordial Density Perturbation}, Cambridge University Press.  \\

\noindent  Malik, K. A. (2007) A not so short note on the Klein-Gordon
equation at second order, {\it JCAP} {\bf 03}, 004 (1-12).  \\

\noindent Malik, K. A. and Wands, D. (2009)
Cosmological perturbations, {\it Physics Reports} {\bf 475}, 1-51.\\

\noindent Martin-Garcia, J. M. and Gundlach, C. (2002),
Self-similar spherically symmetric solutions of the massless
Einstein-Vlasov system, {\it Phys. Rev. D} {\bf 65} 084026. \\

\noindent Mukhanov, V. F., Feldman, H. A. and Brandenberger,
R. H. (1992) Theory of cosmological perturbations, {\it Physics Reports} {\bf 215}, 203-333.\\

\noindent  Mukhanov, V. F. (2005) {\it Physical foundations of cosmology},
Cambridge University Press. \\

\noindent Nakamura, K. (2003) Gauge Invariant Variables in
Two-Parameter Nonlinear Perturbations, {\it Prog. Theor. Phys.}
{\bf 110}, 723-755. \\

\noindent Nakamura, K. (2005) Second Order Gauge Invariant
Perturbation Theory, {\it Prog. Theor. Phys.} {\bf 113},
481-511. \\

\noindent Nakamura, K. (2006) Gauge-invariant Formulation  of the
Second-order Cosmological  Perturbations, {\it Phys. Rev. D} {\bf 74},
101301(1-5).\\

\noindent Nakamura, K. (2007) Second Order Gauge Invariant
Cosmological Perturbation Theory, {\it Prog. Theor. Phys.} {\bf 117},
17-74. \\


\noindent Noh, H. and Hwang, J. (2004) Second order
perturbations of the Friedmann world model, {\it Phys. Rev. D} {\bf 69}
104011(1-52). \\

\noindent Plebanski, J. and Krasinski, A. (2006) {\it An introduction
to general relativity and cosmology}, Cambridge University Press. \\

\noindent Stewart, J. M. and Walker, M. (1974) Perturbations of spacetimes in
general relativity, {\it Proc. Roy. Soc. London A} {\bf 341}, 49-74. \\

\noindent Tomita, K. (2005) Relativistic second-order perturbations
of nonzero-$\Lambda$ flat cosmological models and CMB anisotropies,
{\it Phys. Rev. D} {\bf 71}, 083504(1-11). \\

\noindent Tsagas C. G., Challinor A. and Maartens R. (2008)
Relativistic cosmology and large-scale structure, {\it Physics
Reports} {\bf 465}, 61-147. \\

\noindent Uggla, C and Wainwright, J. (2011) Dynamics of
cosmological scalar perturbations,  preprint. \\

\noindent Wainwright, J. and Ellis, G.F.R. (1997) {\it
Dynamical systems in cosmology}, Cambridge University Press. \\

\noindent Wald, R. M. (1984) {\it General Relativity}, The
University of Chicago Press. \\

\noindent Weinberg, S. (2008) {\it Cosmology}, Oxford
University Press, Oxford, UK.\\

\noindent Wiesenfeld, K. (2001). ResourceLetter: ScL-1:
ScalingLaws. {\it Amer. J. Phys.} {\bf 69} 938-942.

\end{document}